\documentclass[
	amsmath,
	amssymb,
    aps,
    reprint,
    pre,
    showkeys,
]{revtex4-1}

\usepackage{array}

\usepackage[utf8]{inputenc}
\usepackage{amsthm}
\usepackage{hyperref}
\usepackage{graphicx}
\usepackage[font=small,justification=centerlast,format=plain]{caption}
\usepackage{subcaption}
\usepackage{xcolor}
\usepackage{dsfont}
\usepackage{booktabs}
\usepackage{bm}
\usepackage[capitalise]{cleveref}
\usepackage{tikz}
\usepackage{pgf} 
\usepackage{color}

\begin{document}

\title{Gromov Centrality: A Multi-Scale Measure of Network Centrality Using Triangle Inequality Excess}
\author{Shazia'Ayn Babul}
\email{shazia.babul@maths.ox.ac.uk}
\affiliation{Mathematical Institute, University of Oxford, Oxford, UK}
\author{Karel Devriendt}
\email{karel.devriendt@maths.ox.ac.uk}
\affiliation{Mathematical Institute, University of Oxford, Oxford, UK}
\affiliation{Turing Institute, London, UK}
\author{Renaud Lambiotte}
\email{renaud.lambiotte@maths.ox.ac.uk}
\affiliation{Mathematical Institute, University of Oxford, Oxford, UK}
\affiliation{Turing Institute, London, UK}

\date{\today}

\begin{abstract}
Centrality measures quantify the importance of a node in a network based on different geometric or diffusive properties, and focus on different scales. Here, we adopt a geometrical viewpoint to define a multi-scale centrality in networks. Given a metric distance between the nodes, we measure the centrality of a node by its tendency to be close to geodesics between nodes in its neighborhood, via the concept of  triangle inequality excess. Depending on the size of the neighborhood, the resulting Gromov centrality defines the importance of a node at different scales in the graph, and recovers as limits well-known concept such as the clustering coefficient and closeness centrality. We argue that Gromov centrality is affected by  the geometric and boundary constraints of the network, and illustrate how it can help distinguish different types of nodes in random geometric graphs and empirical transportation networks.
\end{abstract}


\maketitle

\section{Introduction}
Network science provides a powerful framework to represent complex systems, by identifying their components as nodes and their interactions as pair-wise edges \cite{newman2010networks}.
Networks are capable of describing systems from a vast array of disciplines, including epidemiology, the social sciences and urban design, and a variety of methods have been designed to extract information from their myriad of connections. Among those, the identification of important, central nodes in a network is particularly critical in different contexts, allowing us to determine, for example, which individuals to vaccinate to hinder a contagion dynamics \cite{pastor2002immunization}, who to contact to trigger a marketing campaign \cite{chen2009efficient}, or which subway stations are essential for connecting areas of a city \cite{crucitti2006centrality}. Various notions of centrality have been proposed to quantitatively compare the importance of nodes in a given network, each based on a different notion of importance \cite{bovet2021centralities}. For example, betweenness centrality gives greater importance to nodes that stand on many of the shortest paths between other nodes.  In a social network where information is flowing between people,  a member with high betweenness centrality is highly influential, best positioned for passing information between other people. Other measures of centrality have been built on the notion of shortest paths, such as closeness centrality, or on properties of the nodes, such as their degree or the number of (not necessarily shortest) walks going through them.

Most centrality measures focus on either the local or global properties of the node within a network. At the local level, a good example is degree centrality, defining the importance of a node solely based on the size of its direct neighbourhood. Relatedly, the clustering coefficient is a local measure of cohesion, and defined as the fraction of closed triangles that exist between the neighbours of a given node. At the global level, closeness centrality measures the mean distance between a node and all other nodes in the graph, typically using the shortest path (geodesic) distance. Similarly, betweenness centrality focuses on shortest paths between any pair of nodes in the network. Yet, it has been argued in the literature that revealing the centrality of a node, or other patterns, at intermediate scales is sometimes more instructive or more efficient \cite{arnaudon2020scale,peel2018multiscale}. Take a large social network like Facebook as a motivating example. Measuring the betweenness of a node is neither practical, as calculating the shortest path between all pairs of nodes is computationally prohibitive, nor insightful, as the shortest paths between nodes at a far away distance do not have a clear meaning. For this reason, different approaches have been proposed to capture centrality within a given neighbourhood of a node or within its community. Multi-scale centrality measures include the $k$-betweenness measure  \cite{borgatti2006graph}, which was used to reveal scale-dependent structures in transportation road networks \cite{yamaoka2021local}.  A localised closeness centrality was also proposed to  analyse street networks \cite{porta2012street}.  Relatedly, classical measures like Katz centrality or Pagerank give more or less importance to longer walks, depending on a parameter \cite{langville2011google}. In relation with community detection, these ideas also appear when defining the importance of a node inside or between communities for instance \cite{guimera2005functional}. Introduced more recently, the diffusion  multi-scale centrality counts  triangle inequality violations computed over the set of nodes reachable within a certain dynamical timescale $\tau$ \cite{arnaudon2020scale}. 

In this article, we adopt a geometrical viewpoint to define a multi-scale centrality in networks and, more generally, in metric spaces. After choosing a metric distance on the network, e.g. the shortest path distance, we define the Gromov centrality of a target node by considering all pairs of nodes within a given distance of this target, and by measuring the size of the triangle inequality for each corresponding triplet. As we show, at the local level, Gromov centrality depends on the proportion of triangles formed by the neighbours of a node, and a central node exhibits a star-like structure locally; geometrically, this relates to discrete curvature where the presence (absence) of triangles around a node is indicative of a positive (negative) curvature \cite{jost2013Ollivier, eckmann2002curvature}. At the global level, in contrast, Gromov centrality is equivalent to closeness, and a node with a high Gromov centrality can essentially be understood as being at the ``center of mass" of the whole network. In addition to providing a novel interpretation for these two quantities in terms of triangle inequality, Gromov centrality shows that they can be seen as two extremes of a family of centrality measures, opening the possibility to define centrality at intermediate scales. Through a series of illustrations and numerical experiments, we  show that the measure is sensitive to the geometric and boundary constraints of the network, as well as to the connectivity of the node.

This article is organized as follows. In Section II, we give the definition of our centrality measure, and discuss its relationship to other common measures of node centrality. We then examine the relationship between the structure of the graph and our Gromov centrality measure, across various scales, using synthetic and empirical graphs. In Section III, we propose an application of our measure; a method of clustering nodes based on their relative structural importance across various scales, and evaluate the method using both random geometric graphs and empirical transportation networks. 

\section{Multi-Scale Gromov Centrality}
\subsection{Preliminary Definitions} 

\begin{figure}[]
\centering
\includegraphics[scale=0.3]{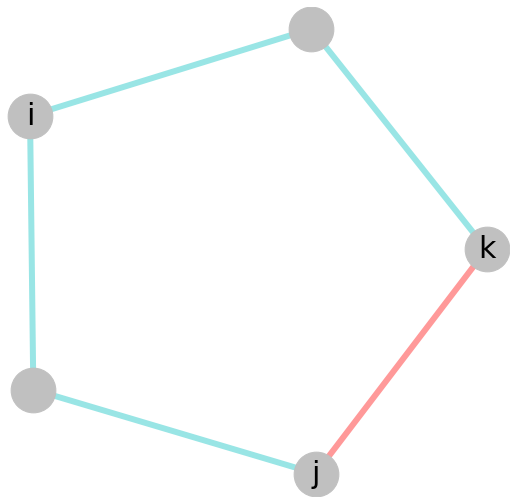}
\caption{In this ring network, if you consider the shortest path between nodes $j$ and $k$, imposing to go through node $i$ significantly increases their distance, and $\Delta_i(j,k)=-3 < 0$. If the red edge is removed from the graph, node $i$ is now on the geodesic between these nodes, and we get $\Delta_i(j,k)= 0$.}
\label{fig:toynet}
\end{figure}

We consider an undirected network composed of $N$ nodes, denoted by the set $V$, described by its symmetric adjacency matrix $A$. We will consider unweighted networks for the sake of simplicity, even if most of the results can be generalised to the weighted case, so that each entry $A_{ij}=1$ ($A_{ij}=0$) encodes the presence (absence) of an edge between nodes $i$ and $j$. The degree of node $i$ is denoted by $k_i = \sum_{j} A_{ij}$.  There exist different ways to define  a metric distance on the nodes of a graph, i.e. a function $d:V\times V\rightarrow [0,\infty)$ that is 
symmetric $d(i,j) = d(j,i)$,  such that $d(i,j)=0$ if and only if $i=j$, and that satisfies the triangle inequality. Some popular examples are the effective resistance and the shortest path distance. In this paper, we will always consider the latter, unless stated otherwise. The triangle inequality can be formulated as the condition
\begin{equation}
\label{eq1}
\Delta_i(j,k) \triangleq d(j,k) - d(i,j) - d(i,k) \leq 0
\end{equation}
for any triplet of nodes $i,j,k$, which intuitively means that the distance between two nodes $j$ and $k$ is always smaller or equal to the length of the shortest path between $j$ and $k$ that passes through an intermediary node $i$. For a target node $i$, and two nodes $j$ and $k$, we call the value of $\Delta_i(j,k)$ the triangle inequality excess between $j$ and $k$ at $i$.
In general metric spaces, the quantity $-\tfrac{1}{2}\Delta_i(j,k)$ is also known as the \emph{Gromov product}, named after Mikhail Gromov who used triangle excesses in his definition of $\delta$-hyperbolic metric spaces~\cite{Gromov1987}. 
$\Delta_i(j,k)$ is equal to zero if and only if node $i$ lies on a geodesic between nodes $j$ and $k$ in the network. Very negative values mean instead that going through node $i$ induces a long detour on the way between $j$ and $k$. As an illustration, consider Figure \ref{fig:toynet}, where the red path is the geodesic between $j$ and $k$, while the blue path is the shortest path going  through  node $i$. In that case, $\Delta_i(j,k)$ would be negative and node $i$ would not play an important role in connecting $j$ and $k$. If the red edge was to be removed from the graph, in contrast,  node $i$ would now lie on the geodesic between $j$ and $k$, and $\Delta_i(j,k)=0$.  

The Gromov centrality of a node $i$  is constructed by considering the triangle excess between all pairs of nodes in a $l$-neighbourhood of node $i$
\begin{equation}
    \Gamma^{l}_{i} = \lbrace j \in V \mid 0 < d(i,j) \leq l\rbrace .
\end{equation}
By construction, when $l$ is the diameter $D$ of the graph the $D$-neighbourhood is the full graph excluding node $i$, and $\vert\Gamma^{l = D}_{i}\vert = N - 1$. Changing the value of $l$ thus allows to tune the size of the neighborhood, from the direct neighborhood to the whole network.
The Gromov centrality of node $i$ at scale $l$ is then defined as 
\begin{equation}
\label{ourmeasure}
    G_{i}^{l} = \frac{1}{\vert T(\Gamma^{l}_{i}) \vert} \sum_{(j,k) \in T(\Gamma^{l}_{i})} \Delta_i(j,k),
\end{equation}
where the sum is taken over the set $T(\Gamma^{l}_{i})$ of pairs of $l$-neighbours  
\begin{equation}
T(\Gamma^{l}_{i}) = \lbrace(j,k)|\; j,k \in \Gamma_{i}^{l}  ~{\rm and }~ j \neq k\rbrace.
\end{equation}
Note that this set does not contain repeats such as $(j,j)$, nor does it contain tuples that include  node $i$, such as $(i,j)$. For $l = D$ the size of the set is then $|T(\Gamma^{D}_{i})| = (N-1)(N-2)$. 

Gromov centrality is thus defined as \emph{the average triangle excess over all pairs of $l$-neighbours of a node $i$.} From the triangle inequality Eq.\eqref{eq1}, it is clear that $G^{l}_{i} \leq 0$, and that the equality holds only if node $i$ is on a geodesic between all pairs. As an example, consider a star graph, with a central node $i$ with its $k_{i}$ neighbours. The central node $i$ has the maximal possible value $G_{i}^{l} = 0$, making it the most central node from our perspective, while all leaf nodes of the star have Gromov centrality $G^{l} = -2$.
Following these arguments, our proposed centrality  characterizes how far a node is from lying `between' all other pairs of nodes in $\Gamma^{l}_{i}$, thus quantifying the importance of the given node at connecting other nodes at scale $l$. Note that Gromov centrality can be calculated directly from the shortest path distance matrix for a given network. 


\subsection{Relation to Other Centrality Measures}
\subsubsection{Arnaudon et al. Diffusion Dynamics Centrality}

In \cite{arnaudon2020scale}, Arnaudon et al.  considered the discrete version of the heat equation, often used to model consensus dynamics or continuous-time random walks, 
$$
\frac{d\mathbf{x}}{dt} = -Q\mathbf{x},
$$
where the Laplacian matrix is defined as  $Q=\operatorname{diag}(\mathbf{k})-A$. They observed that
the local state $x_j(t)$ of a node $j$ in the network can reach a `peak' at a certain time $t^\star>0$. In particular, if we consider a diffusion process starting with all mass at a single node $i$, i.e. with $\mathbf{x}(0)= \mathbf{e}_i$ (for unit vector $\mathbf{e}_i$), then the system states can be found from the matrix exponential $\mathbf{x}(t) = e^{-Qt} \mathbf{e}_i$. One then defines $\tilde{t}^\star(x_i,x_j) = \operatorname{argmax}_{t>0}x_j(t)$, interpreted as the time when the response on $j$ from an impulse on $i$ reaches its maximum, and the \emph{peak-time with time horizon $\tau$} is then defined as 
$$
t^\star_\tau(x_i,x_j) = 
\begin{cases}
\tilde{t}^\star(x_i,x_j) \text{~if this is smaller than $\tau$}\\
\infty \text{~otherwise.}
\end{cases}
$$
It is observed in \cite{arnaudon2020scale} that these peak times contain some information about the `surroundings' of a node in the graph and that $t^\star_\tau(x_i,x_j)$ can be understood as a measure of dissimilarity between the two nodes.
In particular, the measure contains information about whether a node is close to a boundary in the graph or whether it is more central instead. By varying the time horizon $\tau$, this information can be captured over different scales.

To describe the location of a node in the network based on the peak times $t^\star_\tau$, Arnaudon \emph{et al.} propose to count the triangle-inequality violations in which a node is involved. More precisely, for a node $i$ the \emph{multiscale centrality} is defined as
\begin{equation}
\label{arnaudon}
\mathcal{M}^\tau_i = \sum_{j,k\neq i}\mathbf{1}{\lbrace t^\star_\tau(j,i)+t^\star_\tau(i,k)-t^\star_\tau(j,k)\leq 0 \rbrace}
\end{equation}
where $\mathbf{1}\lbrace S\rbrace$ is an indicator function of statement $S$. This multiscale centrality is shown to do very well in identifying central nodes and capturing structural features at different scales \cite{arnaudon2020scale}.

Gromov centrality (\ref{ourmeasure}) is reminiscent of the multiscale centrality (\ref{arnaudon}) as it also uses the triangle inequality and can be assessed over different scales to determine information about the boundary of the graph. However, instead of counting the triangle violations in a `binary way' using the non-linear indicator function, the Gromov centrality counts the net amount by which the triangle inequality deviates from the equality. Furthermore, as the peak-time $t^\star_\tau$ is not an actual metric on the graph it may violate triangle inequality, so that $\mathcal{M}^\tau_i$ counts the number of violations, in contrast with the Gromov centrality, based on the shortest path metric, which by definition cannot violate the inequality. For this reason, the Gromov coefficient is necessarily  nonpositive as it measures the deviation from the triangle inequality. 
\subsubsection{Betweenness Centrality}
Betweenness centrality \cite{freeman1977set} is a classical centrality measure defined in terms of the number of shortest paths that pass through node $i$
\begin{equation}
b_i = \sum_{j,k\neq i}\frac{\#\lbrace j-k \text{ shortest paths through }  i \rbrace}{\#\lbrace j-k \text{ shortest paths}\rbrace}.
\end{equation}
The measure describes, for every pair of nodes, the \emph{fraction of shortest paths} passing through $i$. We notice that the shortest-path distance satisfies
\begin{align*}
&d(j,k)=d(j,i)+d(i,k) 
\\
\Leftrightarrow~ &\exists\text{ a shortest $j-k$ path through $i$},
\end{align*}
and, furthermore, we know that the triangle excess is large when the shortest $j-k$ paths through $i$ are much longer than the shortest $j-k$ paths. 
Gromov centrality indicates how close node $i$ is to lying on a shortest path between all pairs of nodes $j$ and $k$ but, unlike betweenness centrality, it does so
without distinguishing between situations where there are one or several different geodesics between them. Furthermore, the Gromov centrality can be taken at different scales $l$, reminiscent of the local betweenness coefficient introduced in \cite{borgatti2006graph}. 

\begin{figure}
\centering
\includegraphics[scale=0.25]{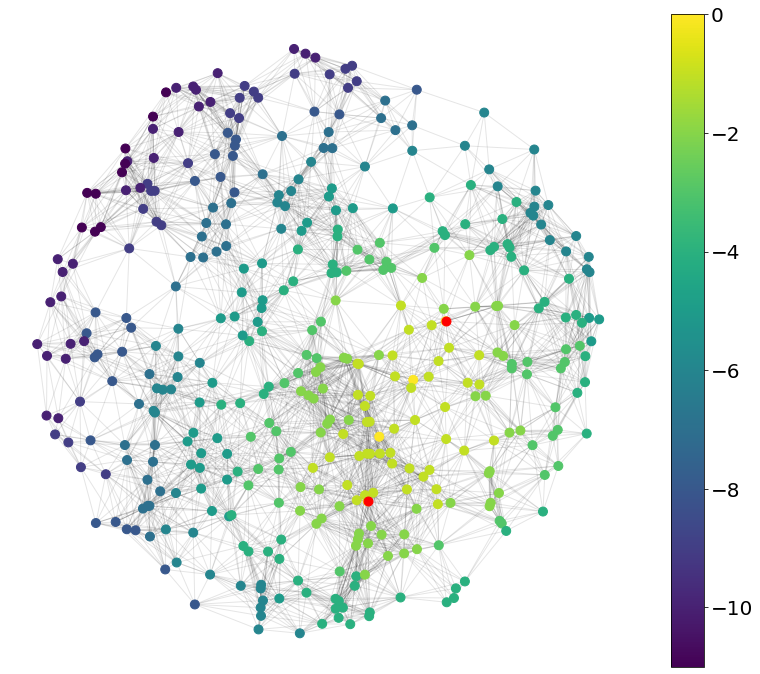}
\caption{Illustration of a non-dense random geometric graph, with parameters $N = 500$ nodes/unit area and $r$ = 0.125. The colours show the triangle excess contribution from nodes $j$ and $k$, indicated in red.}
\label{fig:geonon}
\end{figure}

From this discussion, we conclude that a very negative value of $G_i$ indicates a peripheral node, far from many geodesics, and a small negative value indicates the opposite. 
To further examine the `betweenness' understanding of triangle excess, we consider random geometric graphs. A random geometric graph is characterized by a number of nodes $N$, a domain in a $d$-dimensional space and a characteristic radius $r$. In its simplest setting, one places $N$ nodes uniformly at random within the domain. Two nodes are connected by an edge if the Euclidean distance between them is less than the characteristic radius $r$~\cite{penrose2003random}. Figure \ref{fig:geonon} shows a random geometric graph on the unit disc in $\mathbb{R}^2$. For a fixed pair of nodes $j,k$ indicated in red, every other node $i$ is coloured according to the triangle excess $\Delta_i(j,k)$ from $j$ to $k$ through $i$.
As expected, we observe that the further this node lies from the direct path between nodes $j$ and $k$, the greater the triangle excess will be. 

\begin{figure}
\centering
\includegraphics[scale=0.25]{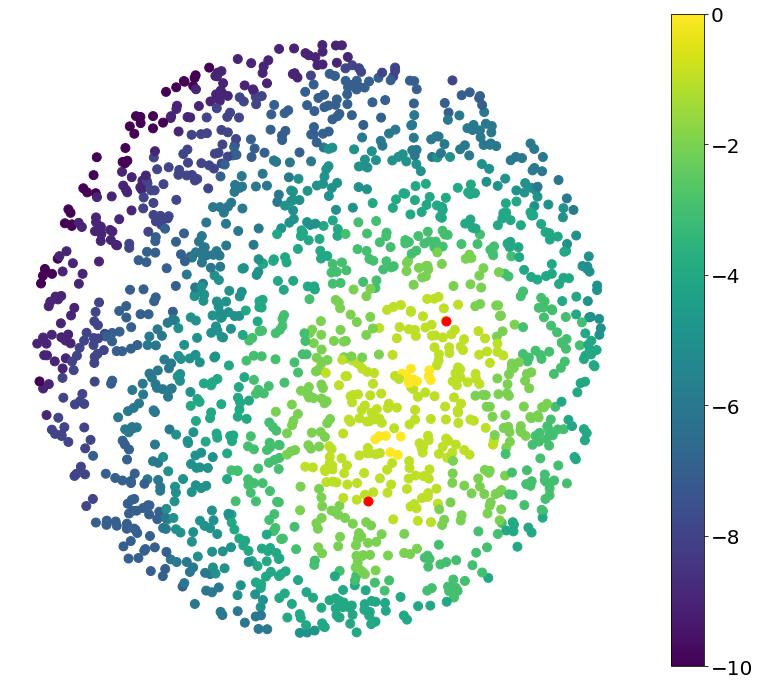}
\caption{Illustration of a dense random geometric graph, with parameters $N = 2000$  nodes/ unit area and $r$ = 0.125. The colours show the triangle excess contribution from nodes $j$ and $k$, indicated in red. The edges have been hidden for the sake of readability.}
\label{fig:geoden}
\end{figure}

For a random geometric graph, the geodesic distance between nodes is bounded from below by $d_{E}(i,j)/r$, where $d_{E}$ is the Euclidean distance, and $r$ is the characteristic radius of the graph. Provided the radius is larger than some critical value, the geodesic distance will also be bounded from above by $K d_{E}(i,j)/r$, where $K$ is some constant.  For a sufficiently dense graph, $K$ is asymptotically equal to 1, and we can approximate the geodesic distance between two nodes as proportional to the Euclidean distance between them \cite{penrose2003random,diaz2016relation}. Figure \ref{fig:geoden} shows the triangle excess contribution of each node for a denser graph. 
The pattern of triangle excess contributions forms ellipses of constant colour, with the red nodes $j$ and $k$ as the two foci. This follows by replacing the distance metric in the triangle excess contribution formula $d(j,k) - d(i,j) - d(i,k)$ by the Euclidean distance, and the fact that all points on an ellipse have the same total distance to the two foci.


\subsubsection{Equivalence with Closeness Centrality at the global scale}

Gromov centrality is  directly related to, and provides an alternative interpretation for, the well-known closeness centrality defined, for each node $i$, as
\begin{equation}
 c_{i} = \frac{N-1}{ \sum_j d(i,j)} =  \frac{1}{\langle d_{i} \rangle}
\end{equation}
where $\langle d_{i} \rangle$ is the average distance from node $i$ to any other node. We consider Gromov centrality at its largest scale, that is when $l$ is equal to 
the diameter $D$ of the graph: 
\begin{equation}
    G_{i}^{D} = \frac{1}{(N-1)(N-2)} \sum_{(j, k) \in T(\Gamma^{D}_{i})} d(j,k) - d(i,j) - d(i,k)
\end{equation}
Starting from the definition of the average distance  
\begin{equation}
    \langle d \rangle = \frac{1}{N(N-1)} \sum_{(i,j) \in T(V)} d(i,j),
\end{equation}
where $T(V)$ is set of all tuples of nodes in the graph, without repeats such as $(i,i)$, we have
\begin{equation}
\sum_{(i,j) \in T(V)} d(i,j)  = \sum_{(j,k) \in T(\Gamma^{D}_{i})} d(j,k) + 2 \sum_{j \neq i} d(i,j)
\end{equation}
and
\begin{equation}
    \sum_{(j,k) \in T(\Gamma^{D}_{i})}  d(i,j) + d(i,k) = 2(N-2) \sum_{j \neq i} d(i,j).
\end{equation}
It follows that Gromov centrality at the global scale can be rewritten as
\begin{equation}
       G_{i}^{D} = \frac{N}{N-2} \langle d \rangle - 2\frac{N-1}{N-2} \langle d_{i} \rangle, 
\end{equation}
which simplifies to
\begin{equation}
       G_{i}^{D} \approx \langle d \rangle - 2\langle d_{i} \rangle = \langle d \rangle - \frac{2}{c_i} 
\end{equation}
 in the large $N$ limit. This expression shows that nodes with a high closeness centrality, have a high (close to zero) value of Gromov centrality. Interestingly, it provides an interpretation of closeness in terms of betweenness, via the notion of triangle inequality, and also allows to generalise closeness to tuneable neighborhoods, via Eq.\eqref{ourmeasure}.
Note that  $G_{i}^{D} $ takes its maximal value of zero for nodes such that $ 2 \langle d_{i} \rangle = \langle d \rangle$.

\subsubsection{Equivalence with Clustering Coefficient at the local scale}

At the smallest scale $l=1$, the Gromov coefficient recovers (minus) the clustering coefficient. The $1$-neighbourhood of a node $i$ satisfies
\begin{equation}
    |\Gamma^{1}_{i}| = |\lbrace j \in V: 0 < d(i,j) \leq 1\rbrace| = k_{i}
\end{equation}
where we recall that $k_{i}$ is the node degree. Consequently, the Gromov centrality measure at this scale is given by
\begin{equation}
    G_{i}^{1} = \frac{1}{k_{i}(k_{i}-1)} \sum_{(j, k) \in T(\Gamma^{1}_{i})} \Delta_i(j,k)
\end{equation}
As with the definition of the clustering coefficient, we require that $k_{i} > 1$. 
The triangle excess is $\Delta_i(j,k)=-1$ if two neighbours $j,k$ of $i$ are connected, and equal to zero if they are not connected. As a result, $G_i^1$ can be written as
\begin{align*}
G_{i}^{1} &= - \frac{1}{k_{i}(k_{i}-1)} \sum_{(j,k) \in T(\Gamma^{1}_{i})} \mathbf{1} \lbrace j,k \text{ are connected}\rbrace \\ \\
&= -  \frac{2}{k_{i}(k_{i}-1)} (\text{number of triangles with node }i)
\end{align*} 
The factor of 2 arises because the set of tuples $T(\Gamma^{1}_{i})$ contains repeats $(j,k)$ and $(k,j)$, which represent the same triangle. The clustering coefficient is defined in \cite{masuda2016guide} as: 
\begin{equation}
    C_{i} = \frac{2}{k_{i}(k_{i} -1)}(\text{number of triangles with node }i)
\end{equation}
It follows that: 
\begin{equation}
G_{i}^{1} = -C_{i}   
\end{equation}
Locally, the Gromov centrality is identically minus the clustering coefficient for each node. 

\subsection{The Effects of Boundary and Scale}
The Gromov centrality contains information about the structure of the graph, across different scales. From the results of the previous section, we know that central nodes will be those with a high value of closeness centrality at the global scale, while they will be those without triangles (that is locally tree-like) at the local scale. In this section, we illustrate the dependency of Gromov centrality on its scale on computer-generated and empirical networks. 

\subsubsection{Random Geometric Graphs}
For a highly homogeneous graph, like a dense random geometric graph, the most important structural effect is the boundary of the graph. Since the density of such a graph is nearly uniform by construction, each node should have the same Gromov centrality at a fixed scale $l$ by spherical symmetry of the $l$-neighbourhood, \textit{provided the graph boundary is not reachable within that scale}. The graph boundary breaks the spherical symmetry, lowering the Gromov centrality value; a node closer to the boundary lies on fewer geodesics between members of its $l$-neighbourhood. The Gromov centrality, then, contains information about the geometric location of a node within the random graph domain. At the smallest scale $l=1$, the Gromov centrality should be uniform across all nodes that are more than geodesic distance 1 from the boundary of the graph. As the scale $l$ increases, a larger fraction of the graph should ``feel" the boundary of the graph. At large scales, the Gromov centrality should reach a maximum at the geometric center of the graph. 
\begin{figure}
\centering
\includegraphics[scale=0.2]{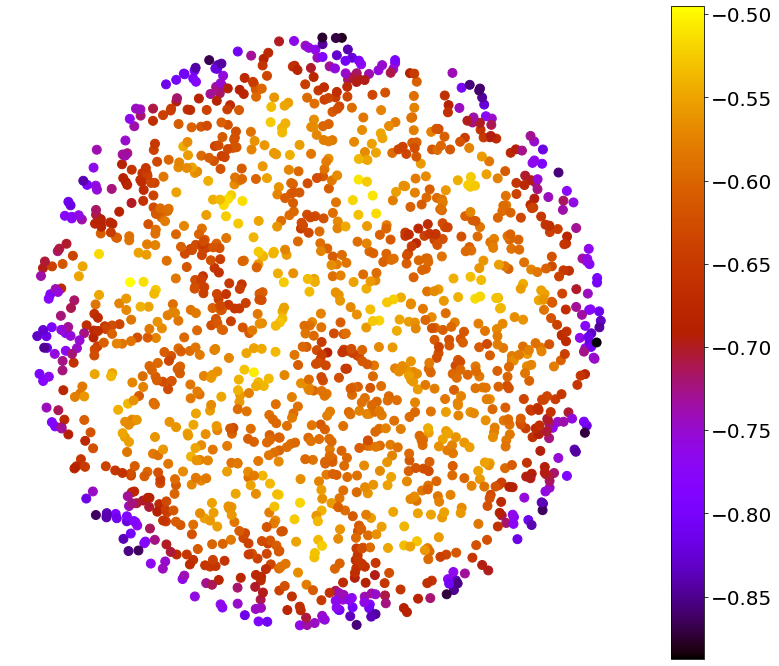}

\includegraphics[scale=0.2]{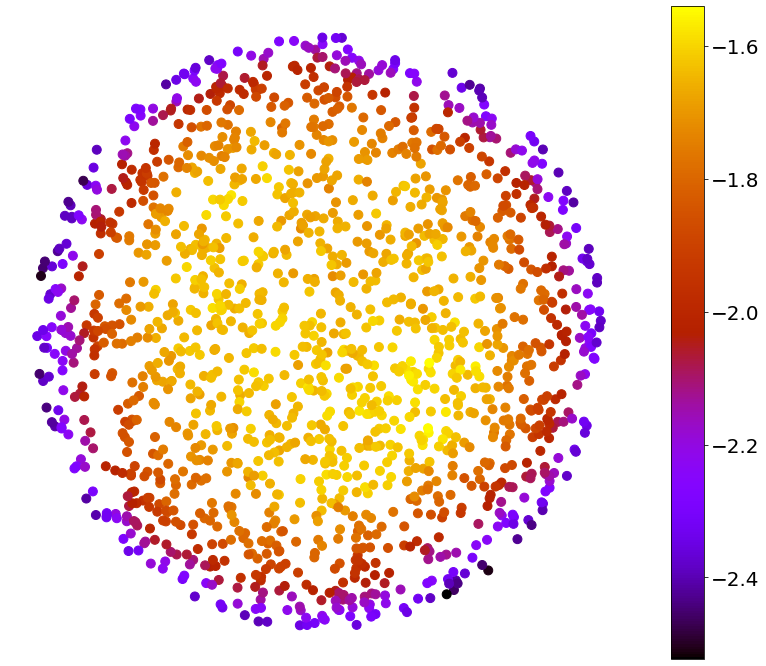}

\includegraphics[scale=0.2]{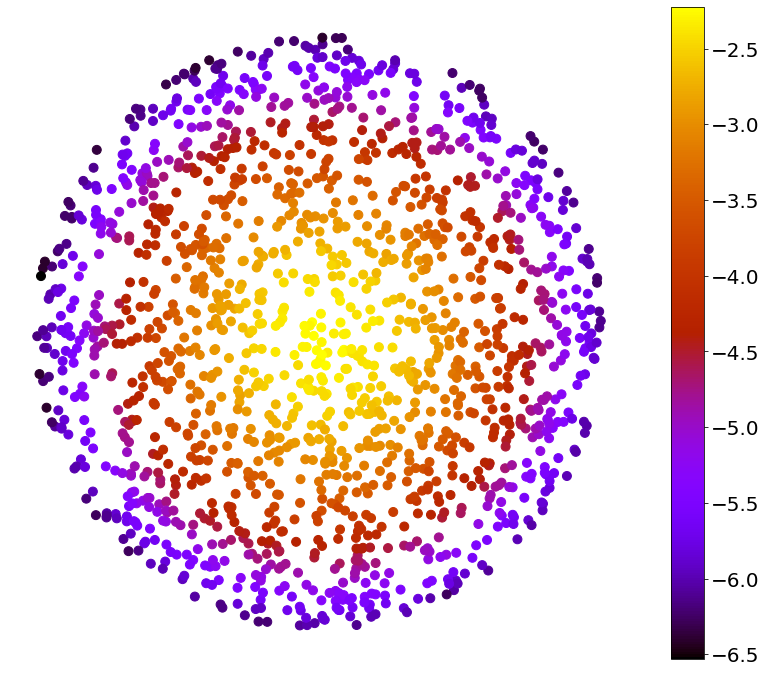}
\caption{Random geometric graphs with density $N$ = 2000 nodes/ unit area. The colours indicates the Gromov centrality value at scales $l=1$, $l=3$ and $l=10$ (from top to bottom).}
\label{fig:l1scale}
\end{figure}

 This is precisely the effect we observe in Figure \ref{fig:l1scale}. At the local scale, nodes lying within $l = 1$ of the boundary form a thin annulus of lower Gromov centrality. The nodes inside this annulus have higher Gromov centrality values, which are not identically uniform due to local inhomogeneities created by random fluctuations in the graph. This annulus grows as $l$ increases. At the larger scale $l = 10$, only nodes at the geometric center of the graph are unaffected by the boundary, and the magnitude of its effect depends continuously on the distance from the center. Also note that fluctuations are less and less pronounced when $l$ increases, which is expected as the size of the neighborhoods increases and more data points are included when estimating the centrality for each node.
 
\subsubsection{Modified Star Graphs}

\begin{figure}
\centering
\includegraphics[scale=0.2]{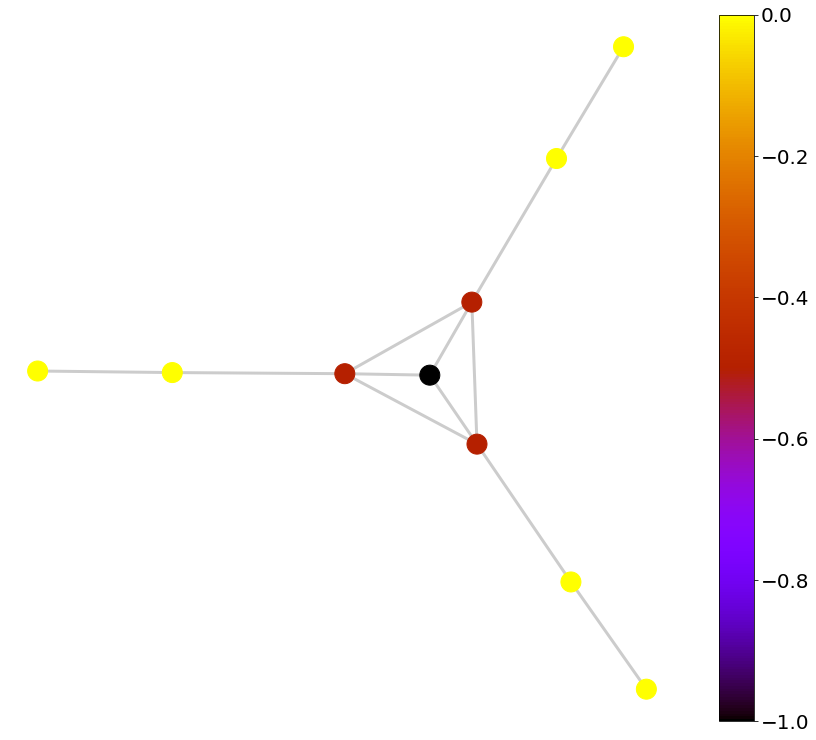}

\includegraphics[scale=0.2]{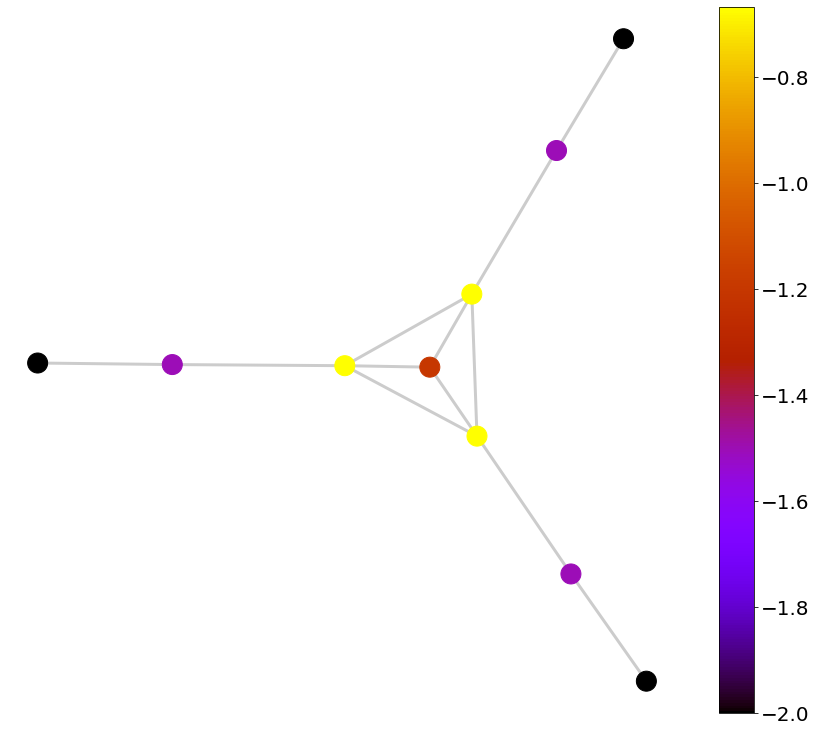}
\caption{Modified star graph. The colours indicate the Gromov centrality value at scale $l=1$ and $l=2$ (from top to bottom).}
\label{fig:modstar1}
\end{figure}

Here, we consider a modified star graph, shown in Figure \ref{fig:modstar1}, to illustrate situations when nodes are central at the local scale, and not at the global scale, and vice versa. First consider the local scale, with $l=1$, where the most central nodes are those with a vanishing clustering coefficient, so that their local neighborhood looks like a star. In this example, these central nodes tend to be on the periphery of the graph. 
When increasing the scale to $l=2$, in contrast, we observe that the centrality of the peripheral nodes drops, and that centrality concentrates on the nodes at the core of the graph, which, despite a high density of triangles, are essential for the connectivity of nodes at the graph periphery, and are thus part of many geodesics. 

\subsubsection{$k$-balanced Tree Graphs}
\begin{figure}
\centering
\includegraphics[scale=0.26]{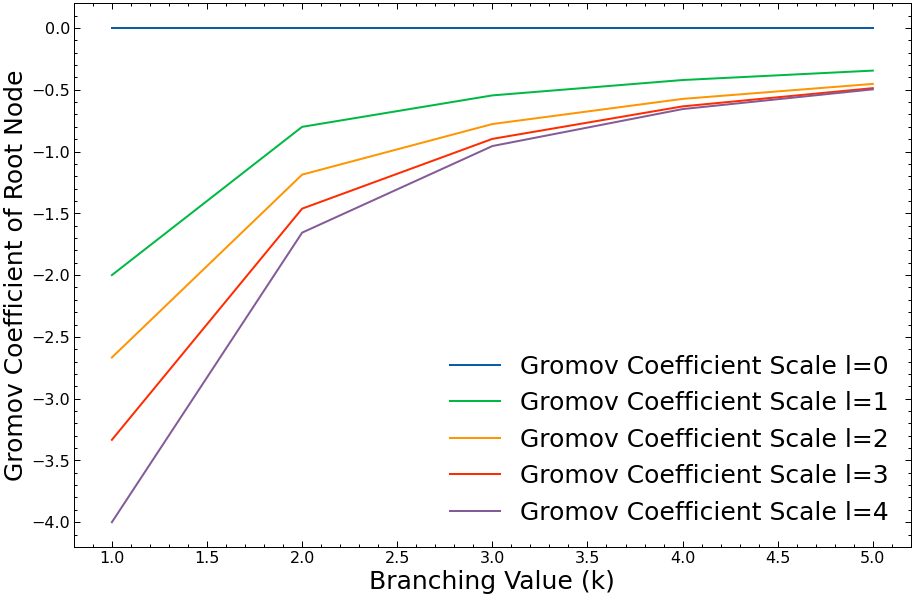}
\caption{Gromov coefficient of root node, for different scales $l$, as a function of the branching factor $k$ of the tree. }
\label{fig:ktree}
\end{figure}

As a next step, we consider the relationship between the Gromov coefficient and nodes that are tree-like, at various scales. $k$-balanced tree graphs are perfectly symmetric graphs of a given height $h$, and branching factor $k$; beginning from a root node with degree $k$, each node branches $k$ times, such that all nodes but the root and the leaves have degree $k + 1$. By construction, there exist no triangles in a tree-graph. In Figure \ref{fig:ktree} we consider the Gromov coefficient of the root node of a $k$-balanced tree, with various branching values, and fixed height $h =5$. The Gromov coefficient at scale $l = 1$, which is minus the clustering coefficient, is trivially 0 for all nodes, by construction of the graph. The Gromov coefficients at other scales converge to 0 as the branching factor increases, due to rapid growth of the neighbourhood size, which is the normalization factor for the coefficient. 
These results indicate that in addition to the importance of cycles, Gromov centrality also depends on the size of the neighborhoods of the nodes, and thus on their degree in particular. 

\subsubsection{Zachary's Karate Club}
To examine the relationship between Gromov centrality and other classical centralities, we consider Zachary's Karate Club graph, a small but well-studied social network \cite{zachary1977information}.
The Karate Club network has 34 nodes, and a diameter of 5. For each of the nodes $i$, we compute the degree, the clustering coefficient, closeness coefficient, betweenness coefficient, and the Gromov centrality for each possible scale  $1 \leq l \leq 5$, and find the Pearson correlations between the measures.  The Pearson correlation was chosen over the Spearman rank correlation because nodes often have the same coefficient values, giving a non-unique ranking. The results are given in  Table I. As expected,  Gromov coefficient is exactly anti-correlated with the clustering coefficient at the smallest scale $l = 1$, while it is highly correlated with the closeness coefficient at the largest scale $l = 5$. 

\begin{table*}[htb]
\label{table1}
\caption{Pearson Correlation between Canonical Centrality Measures and Triangle Violation Centrality at Various Scales (Karate Club)}
\centering
\begin{tabular}{ | m{8em} | m{3em}| m{3em} | m{3em} | m{3em} | m{3em} |} 
\toprule
Canonical Coeff & L =1 & L = 2 & L = 3 & L = 4 & L = 5 \\
\midrule
Degree & 0.52 &  0.96 &  0.74  & 0.74 &  0.71 \\
\midrule
Clustering & -1 &-0.59 &  -0.49 & -0.61 &  -0.61 \\
\midrule
Closeness &  0.61 &  0.88 &  0.94 &  0.99  & 0.98 \\
\midrule
Betweenness & 0.48 &  0.87 &  0.69 &  0.67  & 0.63 \\
\bottomrule
\end{tabular}
\end{table*}

\begin{table*}[htb]
\label{table2}
\caption{Pearson Correlation between Triangle Violation Centralities at Various Scales (Karate Club)}
\centering
\begin{tabular}{ | m{3em} | m{3em}| m{3em} | m{3em} | m{3em} | m{3em} |} 
\toprule
         & L =1 & L = 2 & L = 3 & L = 4 & L = 5 \\
\midrule
L = 1 & 1. & 0.59 & 0.49 & 0.61 & 0.61 \\
\midrule
L = 2 & 0.59 & 1. & 0.82 &  0.86 &  0.83 \\
\midrule
L = 3 &  0.49 &  0.82 &  1. & 0.94 &  0.94 \\
\midrule
L = 4 & 0.61 &  0.86 & 0.94 &  1.& 0.99 \\
\midrule
L = 5 & 0.61 &  0.83 &  0.94 & 0.99 & 1. \\
\bottomrule
\end{tabular}
\end{table*}

Interestingly, we notice that the $l = 2$ scale is highly correlated with both the degree and the betweenness coefficient. The multi-scale centrality introduced in \cite{arnaudon2020scale} applied to the same Karate Club network also demonstrates a high correlation with betweenness at small scales, which the authors attribute to the particular structure of the network. They provide a secondary example of a social network in which betweenness centrality correlates most strongly with their measure at intermediate scales. Consequently, the network-specific structure is highly important for determining the interpretation of the multi-scale dependent centrality measure across scales. 

Table II shows the Pearson correlation between the Gromov centrality measures at different scales. Unsurprisingly, similar scales show a higher correlation. However, we also observe that there is a large change between the $l = 1$ and $l =2$ scales, and that the $l =2$ case actually  correlates more strongly with the global scale than it does with the local one $l = 1$.

\section{Applications}
\subsection{Structural Roles and Node Clustering}
As the previous examples illustrate,  Gromov centrality is particularly adept at capturing the effects of graph structure, and how the ‘centrality’ of a node varies with scale. Here, we build on this insight and propose a method of clustering nodes defined by their role in the network, in terms of the relative structural importance across various scales. Our methodology draws inspiration from the work of \cite{cooper2010role}, in which the authors cluster nodes in directed networks based on the pattern of incoming and outgoing flows across various lengths, and belongs to the broader family of role-detection/node-similarity methods that exploit a variety of scales \cite{blondel2004measure, scholkemper2021local}. Here, we proceed as follows. Each node is assigned a profile vector composed of its (normalized) Gromov coefficients at representative scales, ranging from the local, to intermediate to global. Clustering this matrix results in groupings of nodes that have similar relative geodesic importance across scales. Note that the resulting clusters should not be understood as communities, as they need not be densely connected together, but they instead exhibit similar patterns across scales. In that sense, the resulting groups may be associated to different node roles, which may include disassortative communities and more general block structures \cite{lambiotte2021modularity}. In the following, we evaluate the performance of our methodology on both synthetic random geometric graphs, which are largely homogeneous, and with more heterogeneous empirical transportation networks. 

\subsubsection{Random Geometric Graphs}
We first use our measure to analyze dense random geometric graphs, which have a near-uniform distribution of nodes.  Recall from the previous section, that the Gromov coefficient for such homogeneous graphs is dependent on the chosen scale $l$ and the distance of the node to the boundary of the graph, as in Figure \ref{fig:l1scale}. Thus, we expect that the Gromov coefficient profile should not significantly vary between nodes \textit{at the same distance from the boundary}.  Consequently, the “role” of the node as determined by our proposed clustering method should simply describe the distance from the centre of the graph.

\begin{figure}
\centering
\includegraphics[scale=0.25]{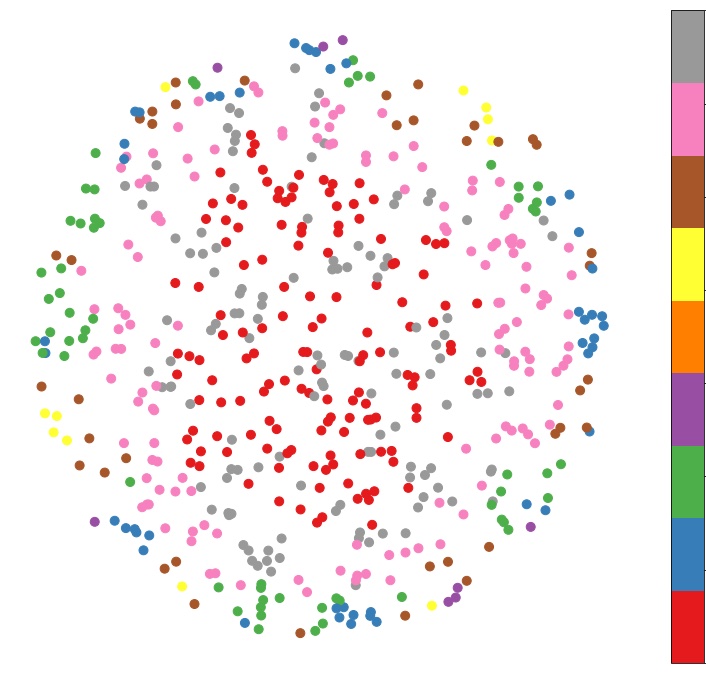}
\caption{Random Geometric Graph, less dense (N = 700/unit area). Node color corresponds to cluster, assigned using 4-vector of l = 1,2,3,4 normalised Gromov coefficients.}
\label{fig:700clust}
\end{figure}

\begin{figure}
\centering
\includegraphics[scale=0.25]{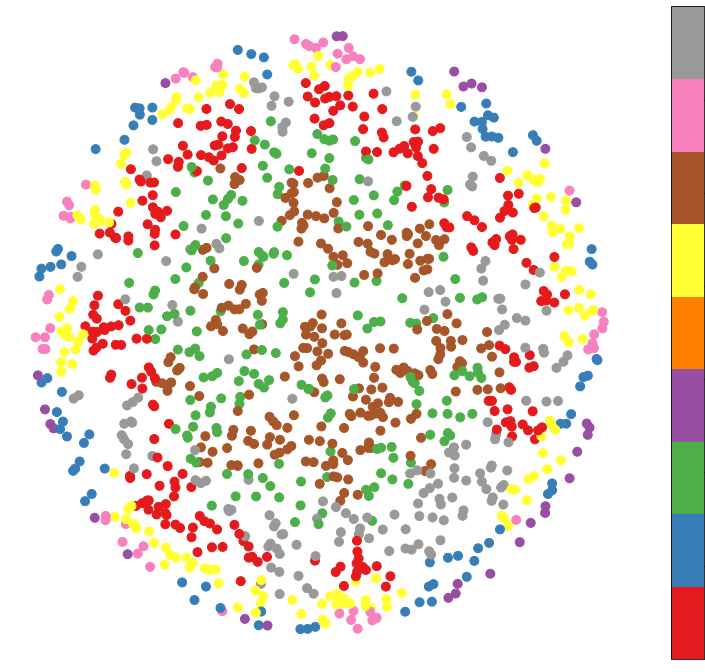}
\caption{Random Geometric Graph, more dense (N = 1000/unit area). Node color corresponds to cluster, assigned using 4-vector of l = 1,2,3,4 normalised Gromov coefficients.}
\label{fig:1000clust}
\end{figure}

The random geometric graphs in Figure \ref{fig:700clust} and Figure \ref{fig:1000clust} have node density of $N =700$/unit area and $N=1000$/unit area, respectively with characteristic radius of $r = 0.125$. Since these graphs have no structural variation, save for local inhomogeneities arising from random fluctuations, we are free to build the node profile from the smallest scale Gromov coefficients for ease of computation. Each node is assigned a 4-vector using the smallest scale Gromov coefficients:
\begin{align}
    v_{i} &= \begin{bmatrix}
           G^{1}_{i} \\
           G^{2}_{i} \\
           G^{3}_{i} \\
           G^{4}_{i}
         \end{bmatrix}
\end{align}
Where each $G^{l}_{i}$ value has been normalized over all coefficients at scale $l$ to lie between -1 and 1. The matrix of Gromov profile vectors is clustered using a spectral clustering algorithm from the scikit-learn package. This algorithm constructs an affinity matrix between the nodes using the radial basis function kernel applied to the Gromov 4-vectors. The affinities are used to create an associated graph, and the clustering is performed using the eigenvectors of the graph Laplacian. \cite{pedregosa2011scikit}  The results are illustrated in Figures \ref{fig:700clust} and \ref{fig:1000clust}, each colour corresponding to a different cluster. As expected, the clusters roughly form concentric circles of nodes with the same radial distance. The radial distance classification is more accurate for the denser graph in Figure \ref{fig:1000clust}, where the geodesic distances between nodes is approximately proportional to the Euclidean distance. 


Varying the chosen scales and number of coefficients used in the vector would change both the number and width of the clusters. Indeed, for these homogeneous random geometric graphs, it is possible to use the largest scale Gromov coefficient (or closeness coefficient) to most accurately estimate the radial distance of a node. However, larger scale Gromov coefficients are more computationally expensive to calculate for dense graphs. Interestingly, as we show in the following analysis, the radial classification is fairly accurate, even for shorter-intermediate scale coefficients.

We analyze how effective the shorter scale Gromov coefficients are at classifying nodes based on radial distance with the following procedure. Using a random geometric graph, we take  two sets of nodes, from the ``inner" area and the ``outer" area of the graph. Using the Gromov coefficient, we cluster the nodes into two groups and examine the fraction of nodes that are correctly identified as inner or outer, as the size of the groups is varied. 

\begin{figure}
\centering
\includegraphics[scale=0.25]{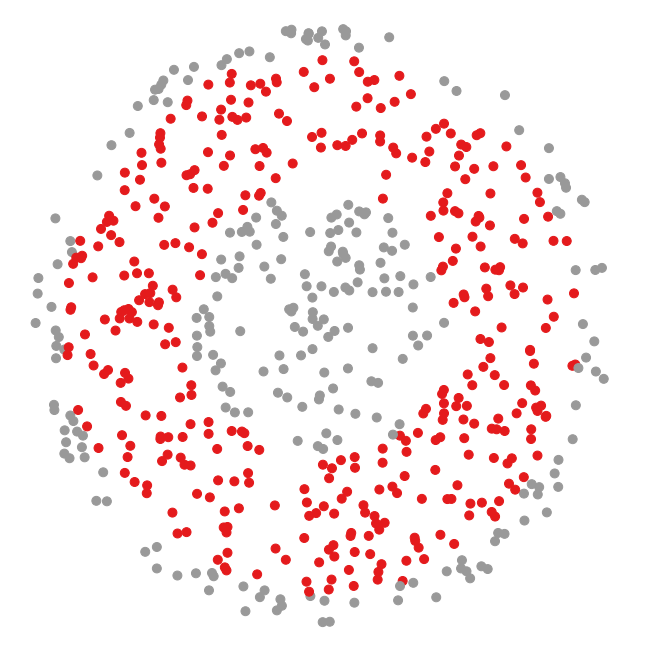}
\caption{Random Geometric Graph. The inner and outer groups are identified in gray, for $x = 0.2$.}
\label{fig:20graph}
\end{figure}

More precisely, we fix a fraction ($x$) of the graph, where $x \leq 0.5$. If $R$ is the radius of the random geometric graph, the inner set is defined as the nodes that fall into the inner $x$ fraction of the graph, or the circle with radius $R\sqrt{x}$.  The outer set of nodes lie in the outer $x$ fraction of the graph, the annulus with radii $R$ and $R\sqrt{1-x}$.  Figure \ref{fig:20graph} identifies the inner and outer sets in grey, for $x= 0.2$. Note that for $x = 0.5$, the union of the inner and outer sets is simply the whole graph. 

\begin{figure}
\centering
\includegraphics[scale=0.25]{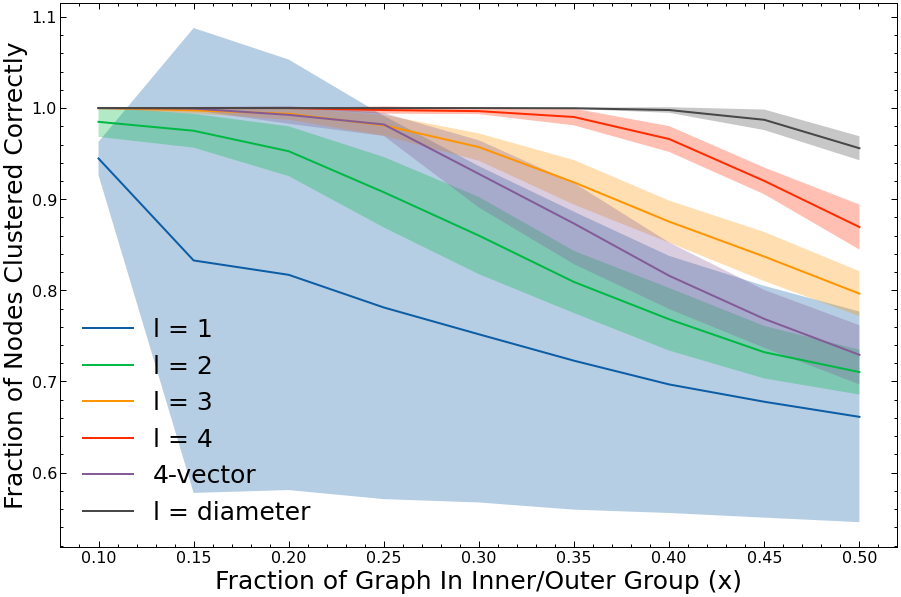}
\caption{Fraction of core nodes correctly identified for different Gromov coefficients as we vary graph percentage, averaged over 10 realizations of random geometric graphs with density N=800/unit area. The shaded area gives the standard error.}
\label{fig:800rgg}
\end{figure}

\begin{figure}
\centering
\includegraphics[scale=0.25]{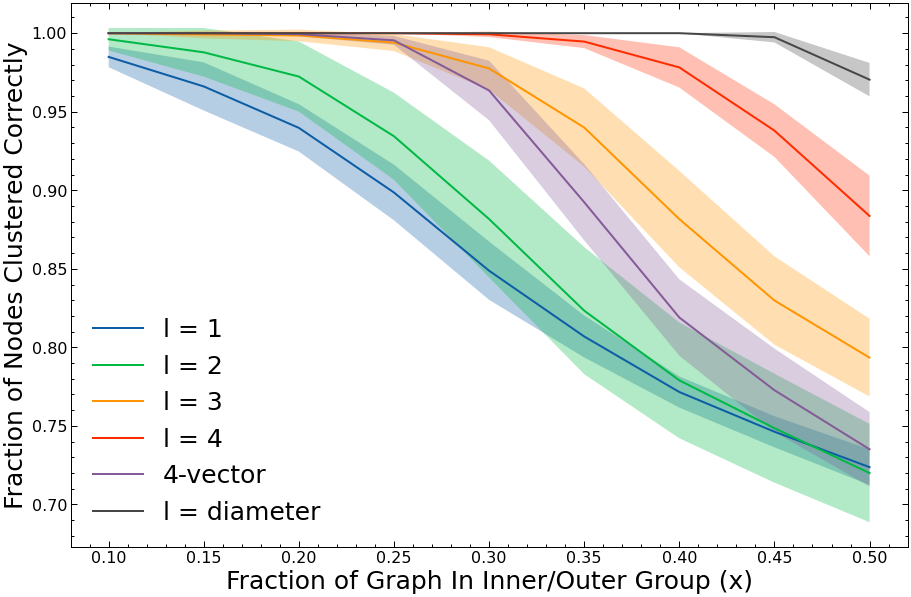}
\caption{Fraction of core nodes correctly identified for different Gromov coefficients as we vary graph percentage, averaged over 10 realizations of random geometric graphs with density N=1500/unit area. The shaded area give the standard error.}
\label{fig:1500rgg}
\end{figure}

Computing the Gromov coefficient for the nodes, we use spectral clustering on the union of the sets, and impose the number of clusters = 2, examining what fraction of nodes are correctly identified as belonging to the inner and outer sets as the $x$ parameter is varied.  As $x$ increases, the classification problem is more difficult, since the variation in the boundary effect between the two sets decreases.  This procedure is carried out for Gromov coefficients of scale $l = 1,2,3,4$, and $l = D$, the diameter of the graph. Finally, the clustering is also performed using the 4-vector profile.  The results are displayed in Figure \ref{fig:800rgg}, for a random geometric graph with density of N = 800/unit area and \ref{fig:1500rgg}, for a random geometric graph with density of N = 1500/unit area. 

As one might expect, the largest scale Gromov coefficients ($l = D$) performs best at clustering nodes based on radial distance; at this scale, the nodes have the most radial variation in the extent to which they can feel the boundary, which is reflected in the value of the coefficient.  However, we also note the effectiveness of the intermediate scale $l=4$ at clustering the two groups, despite the fact that the diameter of the graph is more than double this value. This is an important observation because finding large scale Gromov coefficients is computationally expensive, especially for large graphs. The small-scale coefficients $l =1,2$ perform the worst, and also exhibit the largest errors across iterations, due to local inhomogeneities that occur across the set of graphs.  Consistently across all coefficients, the success of the clustering method decays as the parameter value $x$ increases, increasing the difficulty of the classification problem. Comparing Figures \ref{fig:800rgg} and Figures \ref{fig:1500rgg} shows that the clustering is more successful for the denser graph, across all scales. 

As the density of a random geometric graph increases, the path length is approximately proportional to the Euclidean distance. Thus, increasing the density decreases the local inhomogeneities that introduce fluctuations in the coefficient across nodes at a given radius. As a result, the clustering is more successful, and the standard errors are smaller across all scales. We also note that the success of the 4-vector clustering lies in-between that of the $l=1,2$ and $l=3,4$ scales. This is a consequence of the local fluctuations which introduce smaller scale coefficient variations over nodes at a given radius. This indicates that for such homogeneous graphs, using a Gromov profile actually impedes the clustering, which would be more successful using exclusively the larger scales. 

Dense random geometric graphs are highly homogeneous, and the different Gromov coefficients do not highlight different structures; instead, the different scales simply give finer-grained understandings of the same notion -- distance from the boundary. In the next section, we will pursue our exploration of the method to the case of heterogeneous graphs taken from urban transportation systems. 

\subsubsection{Metro Systems}
Urban transportation systems, and particularly metro systems, have been extensively studied in network science because they are complex networks that exhibit intermediate-scale structure, such as core-periphery architecture \cite{rombach2014core, roth2012long}.  Following the same procedure as for the random geometric graphs, we construct a vector profile of Gromov coefficients across different representative scales to characterize the structural “roles” of the different metro stations,  clustering stations that have similar patterns of relative geodesic importance. The data used in the analysis of the transportation networks was provided by the authors of \cite{roth2012long}.  
The first example is the Paris metro system, which comprises 299 stations, with an average degree of 2.4, and a diameter of 33.  As we vary the chosen scale, we see very different patterns in the Gromov coefficients. We focus mainly on the scales $l = 2, 5, 10$, which give a cross-section of the local, intermediate and global-scale behaviours, while remaining mindful of computation time. Figure \ref{fig:parisl2} in the Appendix displays the results for $l =2$. Note that these graphs are geographically embedded, plotted using the latitude and longitude associated to each station. The leftmost image is coloured according to the Gromov centrality. The middle image is a visual refinement, picking out the ten nodes with the highest valued coefficients. The stations in the rightmost image are coloured according to spectral clustering performed using only the $l =2$ Gromov coefficient.  Examining the resulting pattern created by the Gromov coefficient, and particularly the set of nodes with the highest coefficient values, reveals that this local-scale coefficient gives greater ‘centrality’ importance to stations lying primarily in the outer arrondissements, which serve to connect the periphery to the centre of the city.  The results for $l = 5$, in Figure \ref{fig:parisl5} in the Appendix, illustrate the importance of intermediate-scale Gromov centrality.  The ten nodes with the highest centrality appear to be grouped in pairs, highlighting five areas of the graph, outside the center.  These areas could be understood as intermediate-scale sub-centers.  Although the network has a diameter of 33, the scale $l = 10$ is already sufficiently large to recover the ``closeness" interpretation of large-scale Gromov coefficients, with the additional advantage of a faster computation time.  The $l = 10$ triangle excess coefficient results shown in Figure \ref{fig:parisl10} in the Appendix, illustrates that the centrality peaks at the geometric centre of the graph, or the centre of the city, and fades as it radiates outwards.  The clustering performed using the $l = 10$ coefficient, illustrated in the rightmost image in Figure \ref{fig:parisl10}, groups the nodes into roughly three rudimentary roles, which, as in the random geometric graphs, appear to depend on the geodesic distance from the center. 

\begin{figure}
\centering
\includegraphics[scale=0.18]{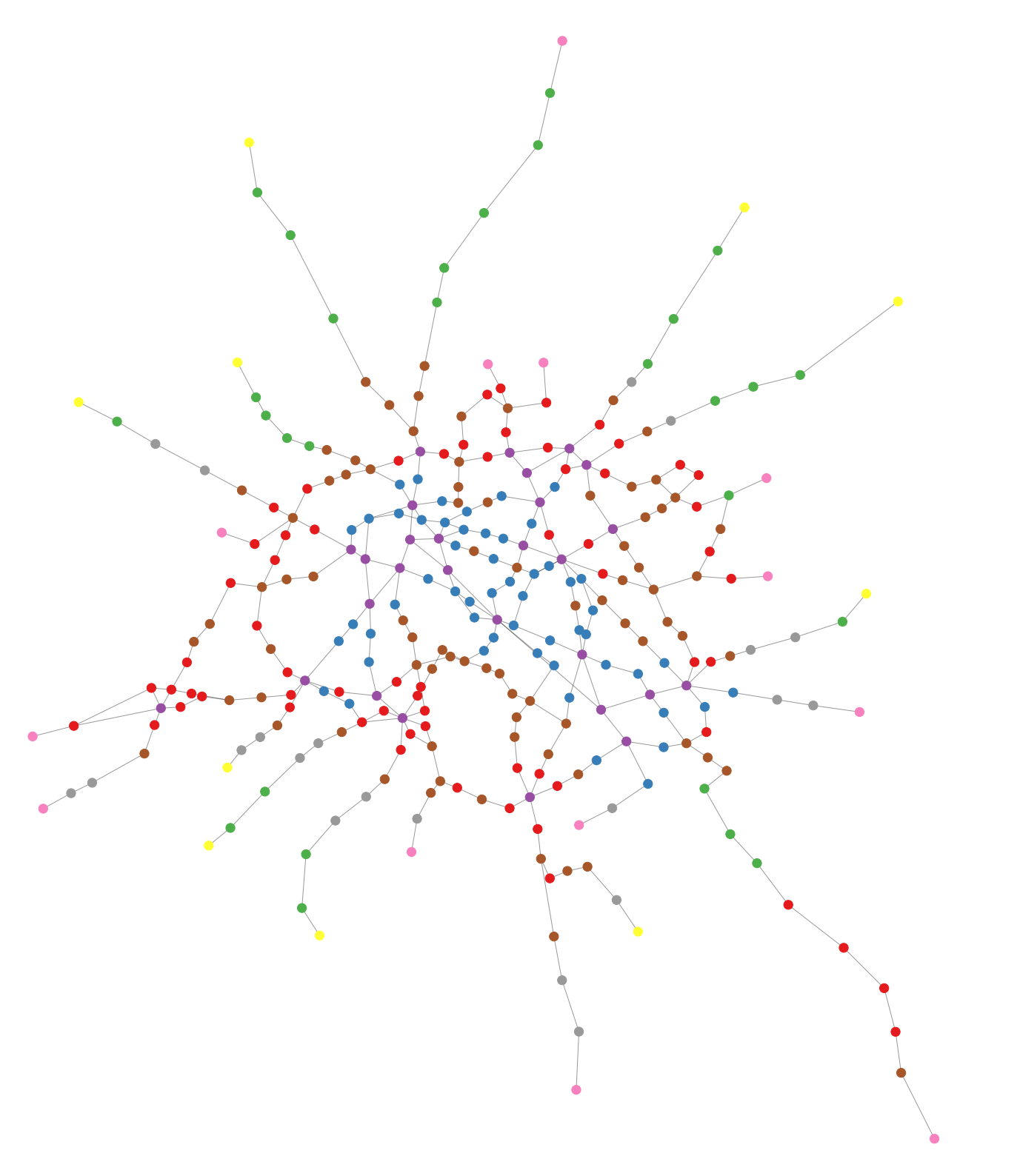}
\caption{Parisian metro system, geographically plotted, clustered using 3-vector of Gromov coefficients $l = 2,5,10$}
\label{fig:3vec}
\end{figure}

Returning to the role-based clustering using the Gromov profile, each station is assigned a 3-vector given by: 
\begin{align}
    v_{i} &= \begin{bmatrix}
           G^{2}_{i} \\
           G^{5}_{i} \\
           G^{10}_{i} \\
         \end{bmatrix},
\end{align}
where each $G^{l}_{i}$ value is again normalized over all coefficients at scale $l$ to lie between -1 and 1, to ensure that the information at each scale is weighted equally. The spectral clustering performed on these vector representations returns 8 clusters.  We note that the clustering method could be refined to produce more consistent results.  The network is shown once again in Figure \ref{fig:3vec}, where each station is coloured by cluster.  

Naively, the colours appear to be laid out according to the distance from the centre. The green, yellow and pink clusters show metro stops leading out into the periphery of the city, with yellow and pink identifying the end of the line.  Brown and red nodes encircle the city centre core, connecting it to the periphery. Blue nodes surround the core even more closely, and finally the purple cluster highlights nodes at the very centre of the city. 

Interestingly, we note that the stations identified by the intermediate $l = 5$ scale as sub-centres, and stations belonging to the main city centre, as identified by the $l = 10$ scale, \textit{are assigned to the same purple cluster}. Figure \ref{fig:3vec} is re-plotted in the Appendix as Figure \ref{fig:3veclabel} with the purple cluster nodes labeled with their station name. Many of these nodes appear to be important train stations, including Montparnasse Bienvenue, Gare du Nord, Gare de Lyon, Gare de l'Est and Saint Lazare. Some of these stations, like Montparnasse Bienvenue, have a relatively low closeness coefficient, but still function as important intermediate hubs, a feature which is indeed captured by our Gromov profile clustering method. 

With a geographical embedding, the Parisian metro system appears to have a core-periphery structure, with metro lines terminating outside the dense city centre. Consequently, the geometric center of the graph is highly indicative of the city center.  As an informative example, we perform a similar analysis on a less centralized transportation network; the New York subway system. The data used in this analysis was also provided by the authors of \cite{roth2012long}. The New York metro system comprises 433 stops, with an average degree of 2.2, and a diameter of 59, and the network is plotted using a geographical embedding. As before, the Gromov coefficients reveal different patterns at different scales. We select the scales $l = 2, 4, 10, 15$, which give a cross-section of local, intermediate and global-scale behaviours.   

\begin{figure}[h!]
\centering
\includegraphics[scale=0.16]{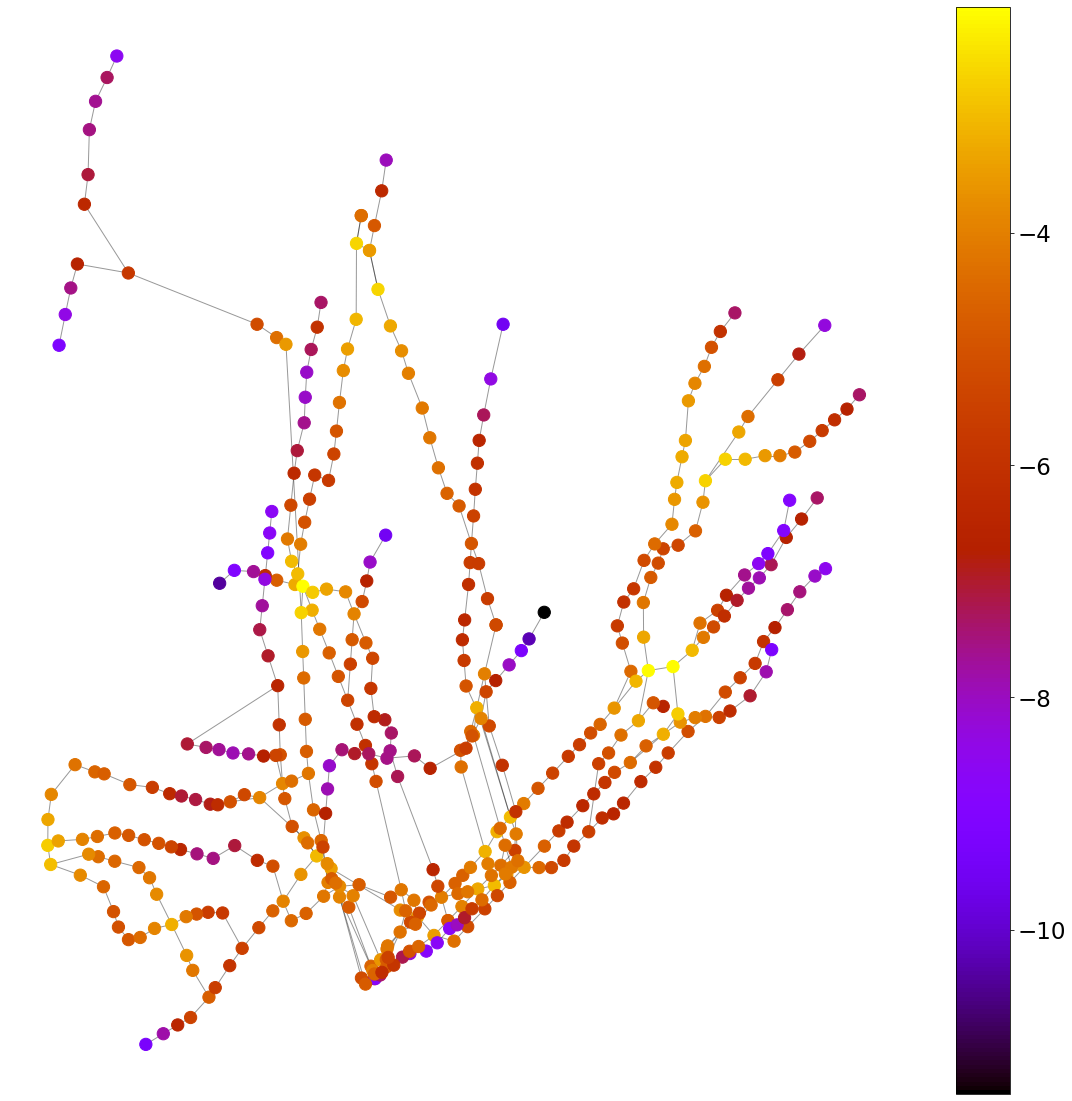}
\includegraphics[scale=0.16]{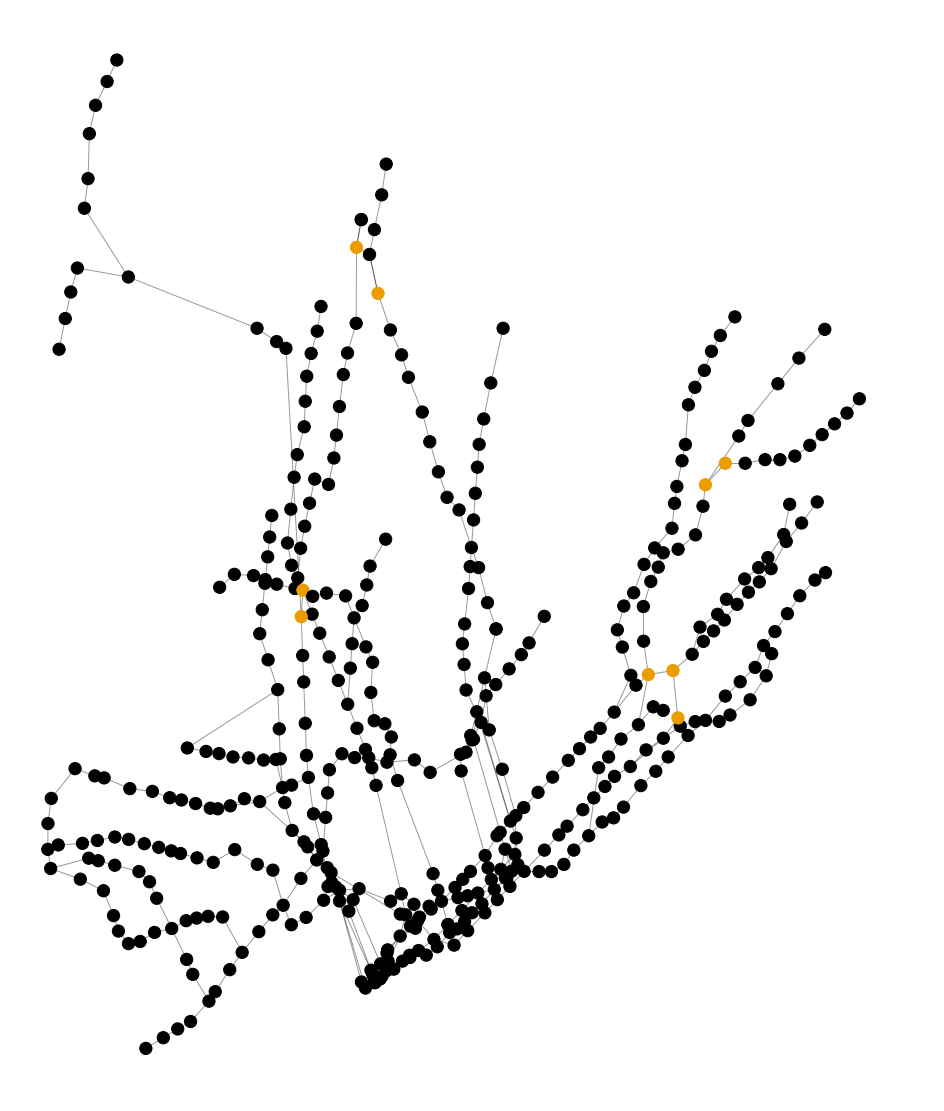}
\includegraphics[scale=0.16]{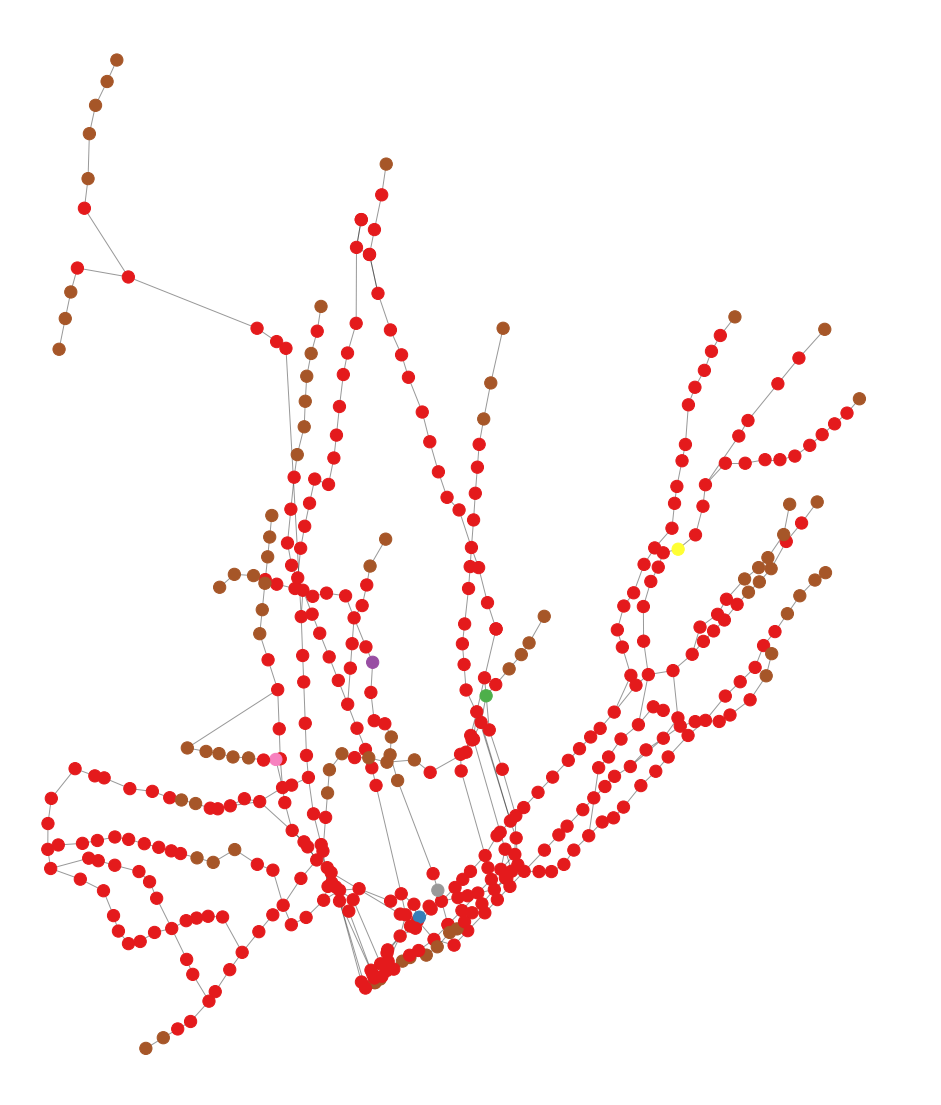}\caption{New York metro system, geographically plotted, results for $l = 10$. The first image shows the value of the $l =10$ Gromov coefficient for the different stops. The second image highlights the 10 stops with the highest $l = 10$ coefficients. The third image shows the results of spectral clustering performed using just the $l = 10$ coefficient. The 10 highlighted stops are: 155th StreetBD, East 180th Street, Rockaway Avenue AC, Bronx Park East, Sutphin Boulevard Archer Avenue JFK Airport, Jamaica Van Wyck, Briarwood Van Wyck Boulevard, 149th Street Grand Concourse, 161st Street Yankee Stadium, Broadway Junction}
\label{fig:nycl10}
\end{figure}

\begin{figure}[h!]
\centering
\includegraphics[scale=0.16]{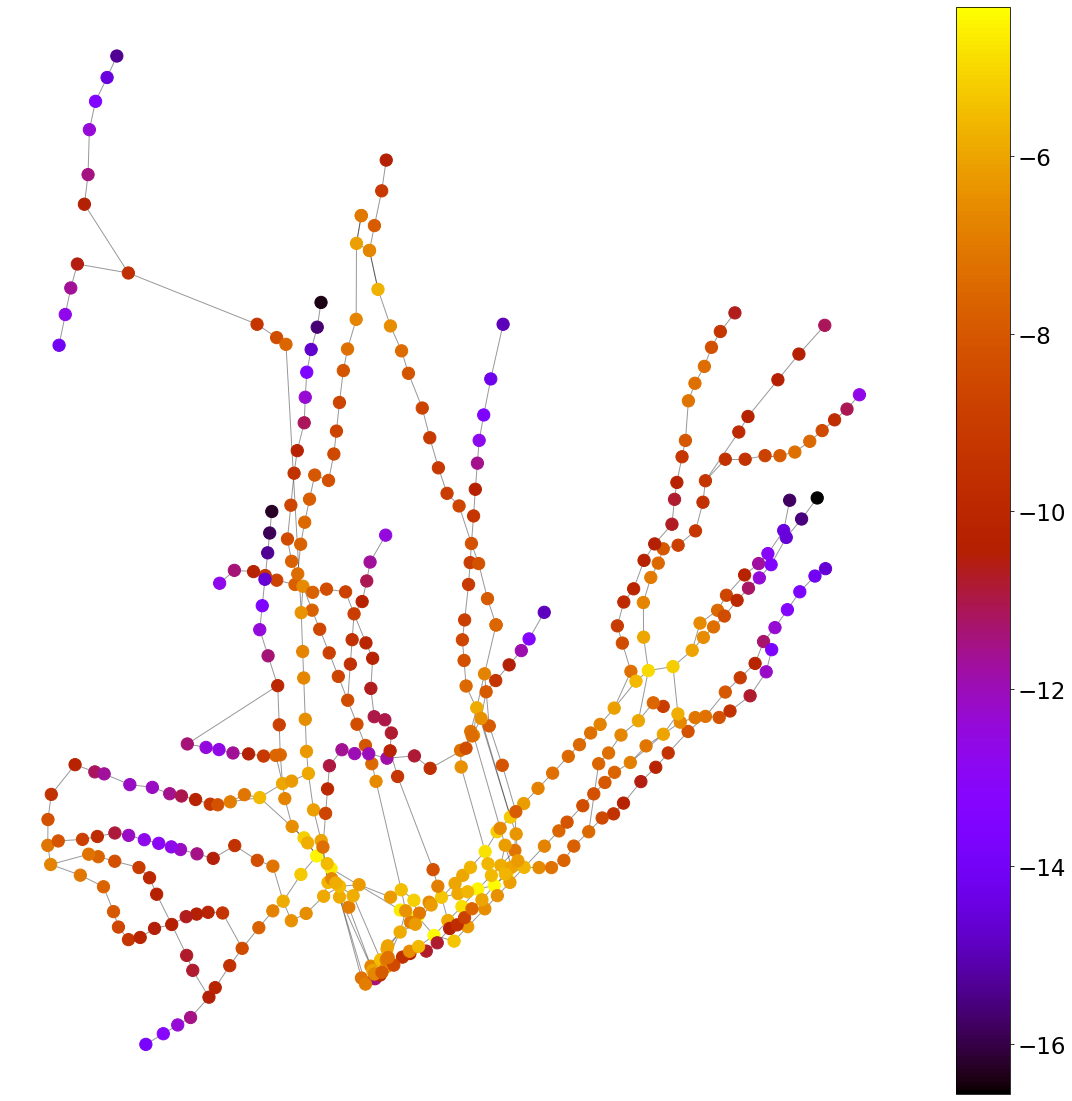}
\includegraphics[scale=0.16]{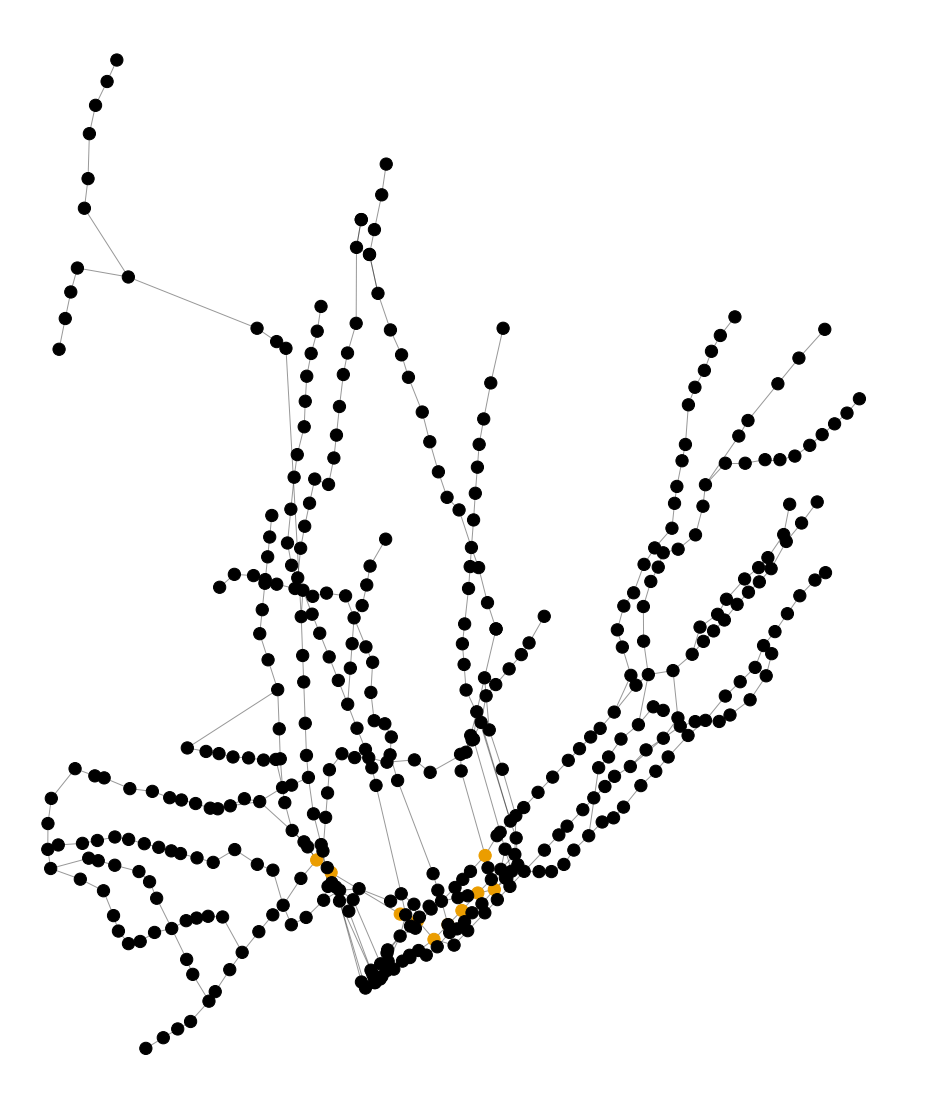}
\includegraphics[scale=0.16]{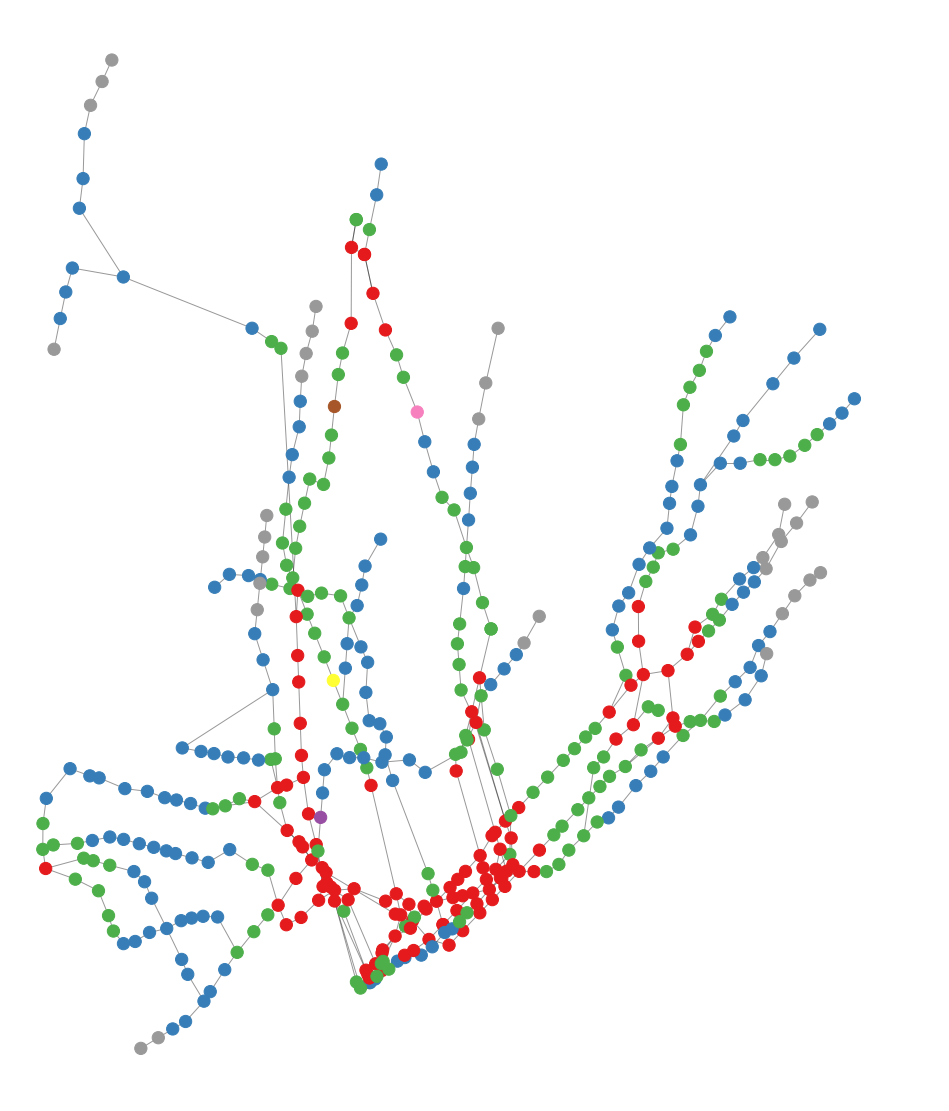}\caption{New York metro system, geographically plotted, results for $l = 15$. The first image shows the value of the $l =15$ Gromov coefficient for the different stops. The second image highlights the 10 stops with the highest $l = 15$ coefficients. The third image shows the results of spectral clustering performed using just the $l = 15$ coefficient. The 10 highlighted stops are: 23rd Street FV, 14th Street FV, Grand Central/42nd Street, DeKalb Avenue X, Grand Street BD, 34th Street Herald Square, Atlantic Avenue/Pacific Street,Times Square/42nd Street Broadway Lafayette Street, West Fourth Street Washington Square.}
\label{fig:nycl15}
\end{figure}

Figures \ref{fig:nycl2} and \ref{fig:nycl4} in the Appendix display the results for the smaller scales, $l=2$ and $l=4$ respectively. We notice that most of the 10 highest coefficient nodes are outside the dense cluster of stations that represents Manhattan. Already, clustering the Gromov coefficients at scale $l=2$ highlights a sub-centre structure in the outer boroughs, which also appears in the 4-vector clustering.  Scaling up to $l = 10$, the highest valued coefficient nodes are still outside Manhattan, but they are now grouped together in what we might identify as sub-centres, as in Figure \ref{fig:nycl10}.  Finally, at scale $l = 15$ in Figure \ref{fig:nycl15}, we recover a closeness-like interpretation of the coefficient; the important nodes are inside Manhattan. Indeed, 8 out of the 10 highest valued nodes identified by the $l = 15$ Gromov coefficient are the same as the 10 identified by the closeness coefficient, despite the graph having a diameter of 59. Intuitively, it is reasonable that long journeys from one outer borough to another are required to pass through Manhattan. The clustering at this scale does identify sub-centres in the outer boroughs, but we observe that most stations inside Manhattan are placed in one large group, indicating that the coefficient at this scale is not fine-grained enough to capture intermediate-scale effects.  

\begin{figure}
\centering
\includegraphics[scale=0.25]{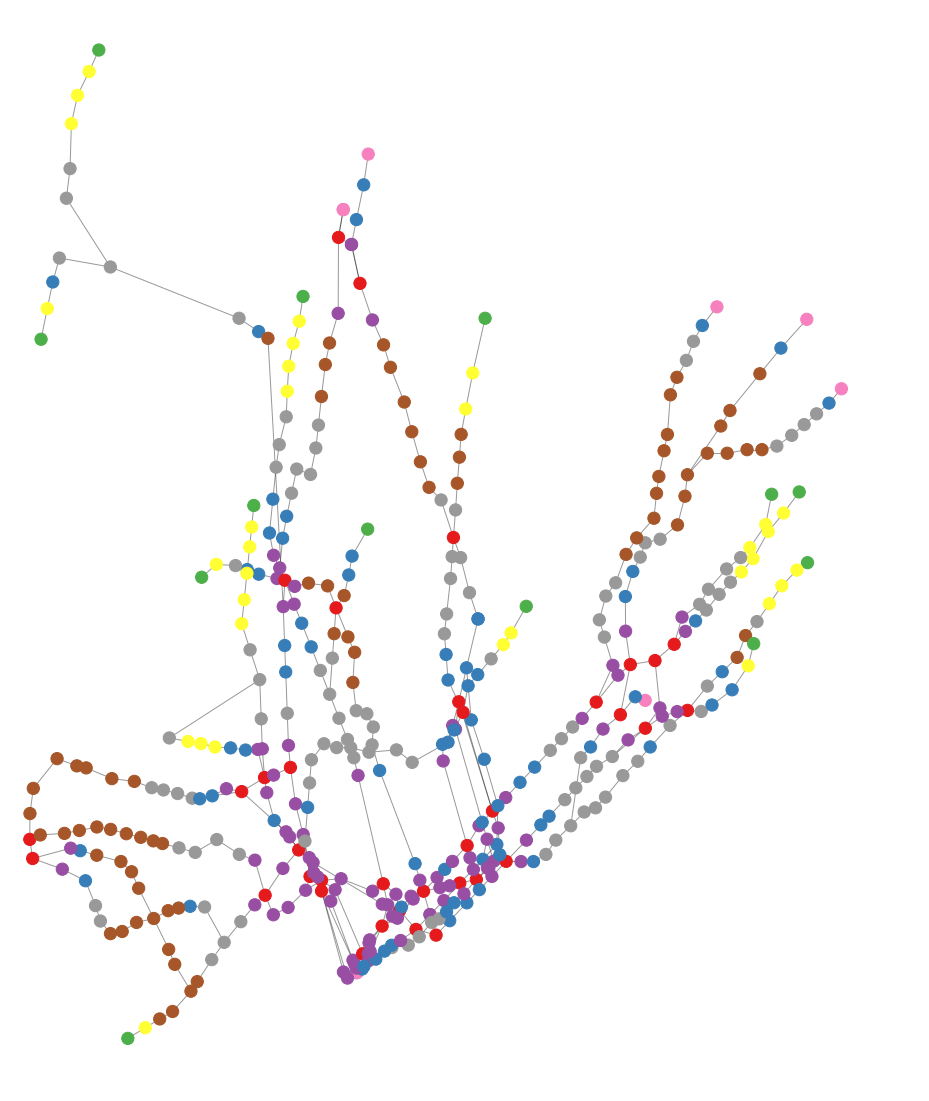}
\caption{New York metro system, geographically plotted, clustered using 4-vector of Gromov coefficients $l = 2,4,10,15$}
\label{fig:4vecnyc}
\end{figure}

Finally, we cluster the metro stops using the vector profile of the selected Gromov coefficients. As before, each node is assigned a 4-vector given by: 
\begin{align}
    v_{i} &= \begin{bmatrix}
           G^{2}_{i} \\
           G^{4}_{i} \\
           G^{10}_{i} \\
           G^{15}_{i} \\
         \end{bmatrix},
\end{align}
where again each $G^{l}_{i}$ value has been individually normalized to lie between -1 and 1. Figure \ref{fig:4vecnyc} shows the geographically embedded New York subway network, where each node has been assigned to a cluster indicated by its colour. We can interpret the red nodes as central core nodes, encircled by layers of purple and blue nodes. The yellow and green identify line terminations, and the stations that function as connectors between sub-centres are identified in grey. While the concentration of central nodes inside Manhattan is not unexpected, we also see the appearance of sub-centres in the outer boroughs; Brooklyn, Queens and the Bronx.  

Paris, which is itself is laid out in concentric circles, exhibits sub-centres that are within the rings of the city, as illustrated in Figure \ref{fig:3vec}. The New York subway on the other hand, as Figure \ref{fig:4vecnyc} depicts, has a more complicated configuration, perhaps as a result of the  bodies of water that separate the different boroughs. The distribution of sub-centres could indicate that New York is an example of a ``polycentric" city, in the sense of \cite{roth2011structure}, where the authors use commuter flow to examine the hierarchical structure patterns that emerge at different scales, identifying polycentric cities as ones that have ``nested urban movements" that can be divided into sub-centers. Here, we have analyzed just the stationary metro system network, and it may indeed be interesting to compare the sub-centers identified by the Gromov coefficient on the stationary network, with the commuter flow-based subdivision performed in \cite{roth2011structure}.

\section{Conclusion}
In this paper, we have presented Gromov centrality, a measure of multi-scale node centrality that computes the average triangle excess over all pairs of nodes in a given $l$-neighbourhood. The measure can be taken across various scales, and captures the importance of a given node at connecting other nodes in its vicinity. We show that at the local scale, Gromov centrality is precisely minus the clustering coefficient, and at the global scale, it recovers a closeness interpretation. Using various toy networks, we show that the measure is sensitive to the geometric and boundary constraints of the graph, at various scales. We propose an application of our measure to role-based clustering of nodes, in terms of the relative structural importance across scales. We show that on random geometric graphs, the clustering method can identify the distance of a node from the boundary. In experiments on heterogeneous empirical transportation networks, the clustering method groups together stations that play similar roles in the flow of transportation, and identifies the various sub-centers that appear in polycentric cities. Overall, this work contributes to the increasing efforts to characterise and to exploit network geometry \cite{boguna2021network}.  Future research perspectives include investigating the relationship between Gromov centrality  and dynamical processes on networks, for instance to study how complex diffusion is affected by structures at different scales in the network \cite{o2015mathematical}, and exploring the behaviour of Gromov centrality for other metric distances between nodes, in particular the effective resistance. 

\begin{acknowledgments}
 SB was supported by EPSRC grant EP/W523781/1. KD was supported by The Alan Turing Institute under EPSRC grant EP/N510129/1. The work of RL was supported by EPSRC grants EP/V013068/1 and EP/V03474X/1.  Package from~\cite{SciencePlots} has been used to make plots.
\end{acknowledgments}

\bibliographystyle{apsrev4-1}
\bibliography{references}
\clearpage
\onecolumngrid
\section*{Appendix}
\begin{figure}[h!]
\centering
\includegraphics[scale=0.16]{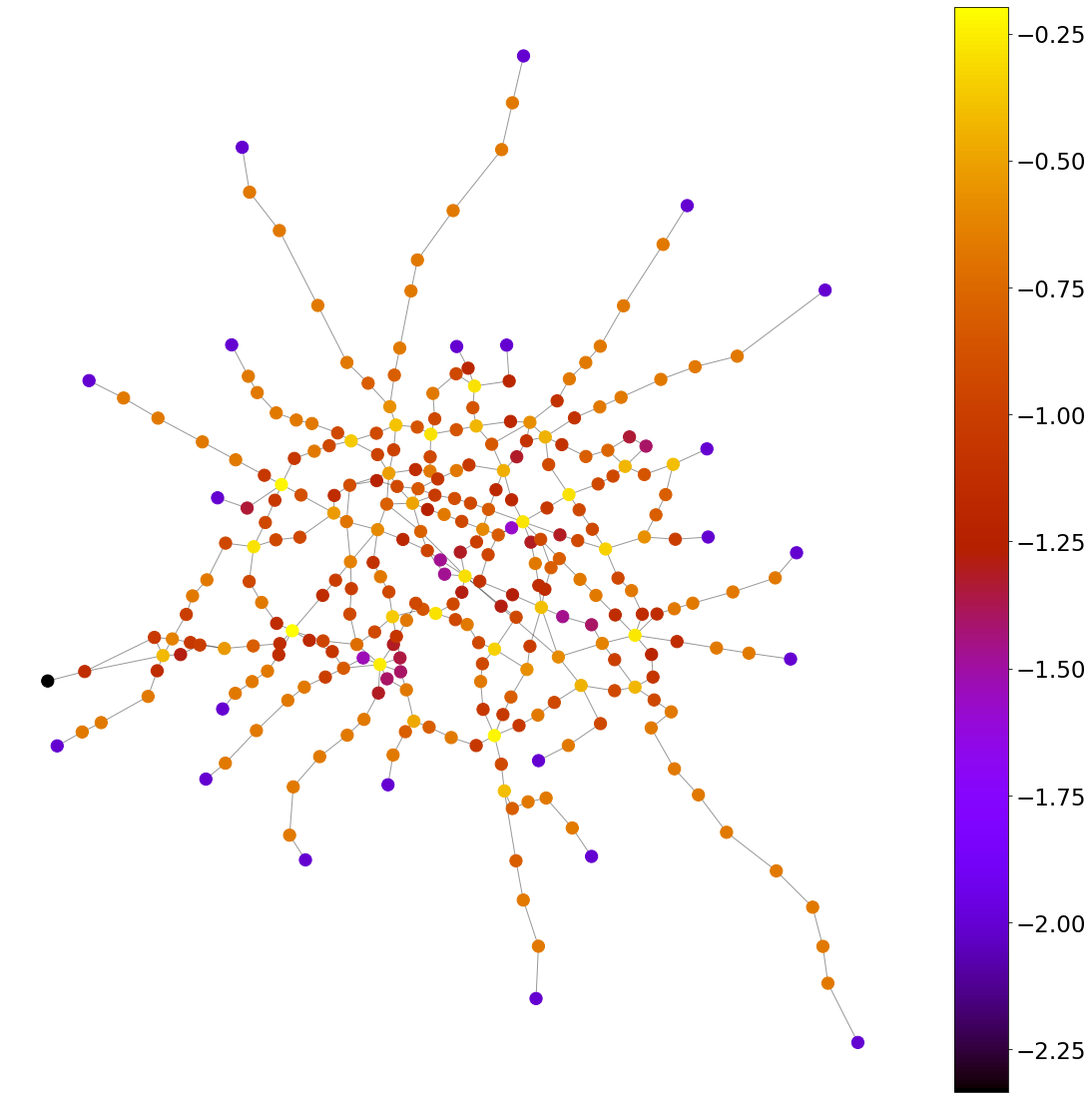}
\includegraphics[scale=0.16]{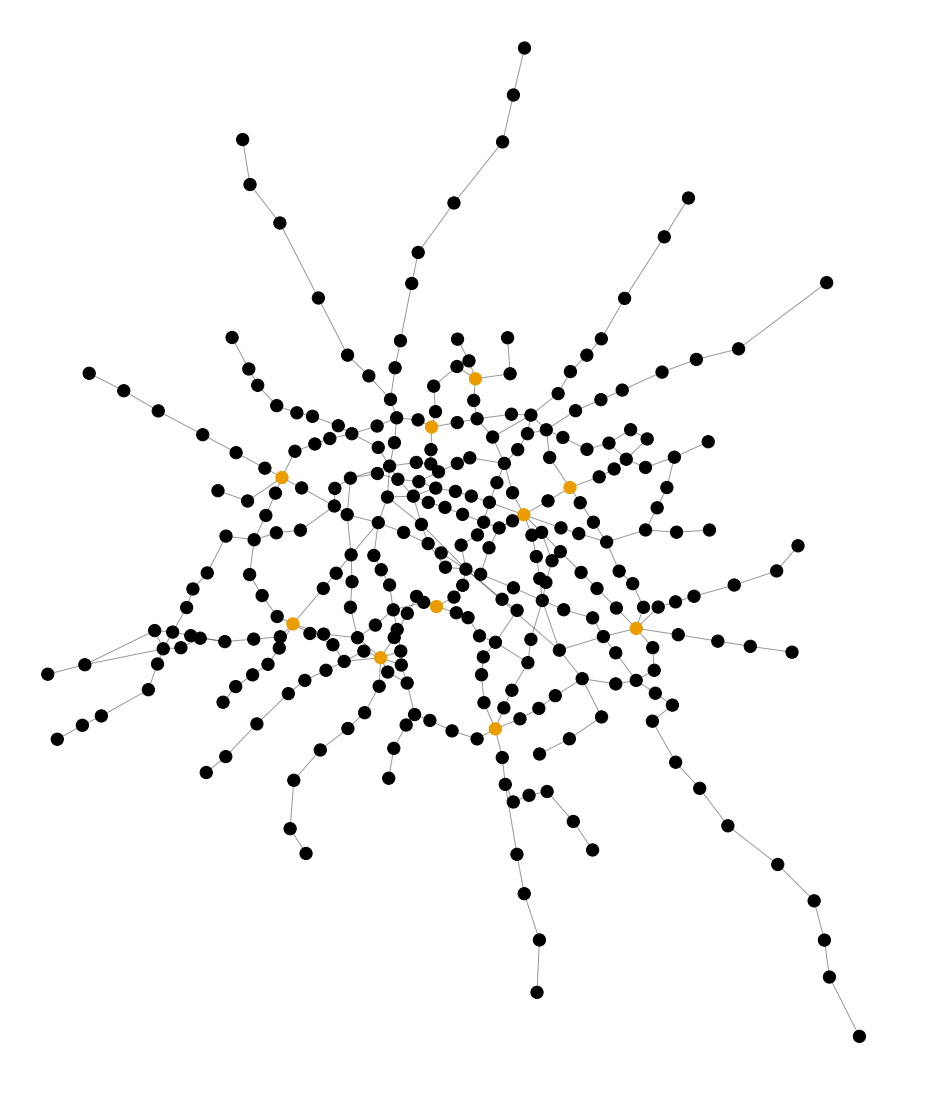}
\includegraphics[scale=0.11]{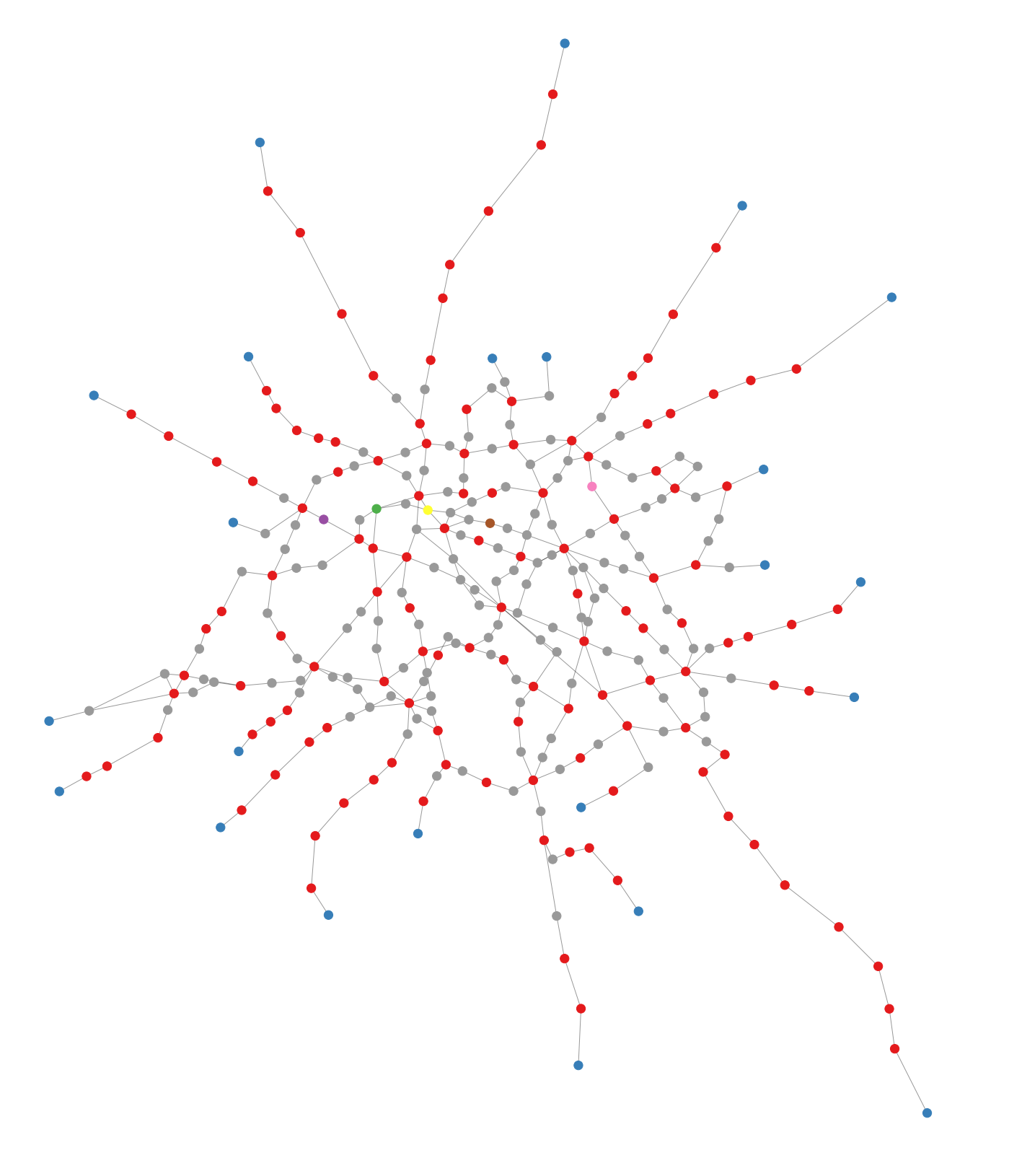}
\caption{Parisian metro system, geographically plotted, results for $l = 2$. The first image shows the value of the $l =2$ Gromov coefficient for the different stops. The second image highlights the 10 stops with the highest $l = 2$ coefficients. The third image shows the results of spectral clustering performed using just the $l = 2$ coefficient. The 10 highlighted stops are: Belleville, Marcadet Poissonniers, Republique, Montparnasse Bienvenue, Odeon, Pigalle, Place d'Italie, Nation, Charles de Gaulle Etoile, La Motte Picquet Grenelle.}
\label{fig:parisl2}
\end{figure}

\begin{figure}[h!]
\centering
\includegraphics[scale=0.16]{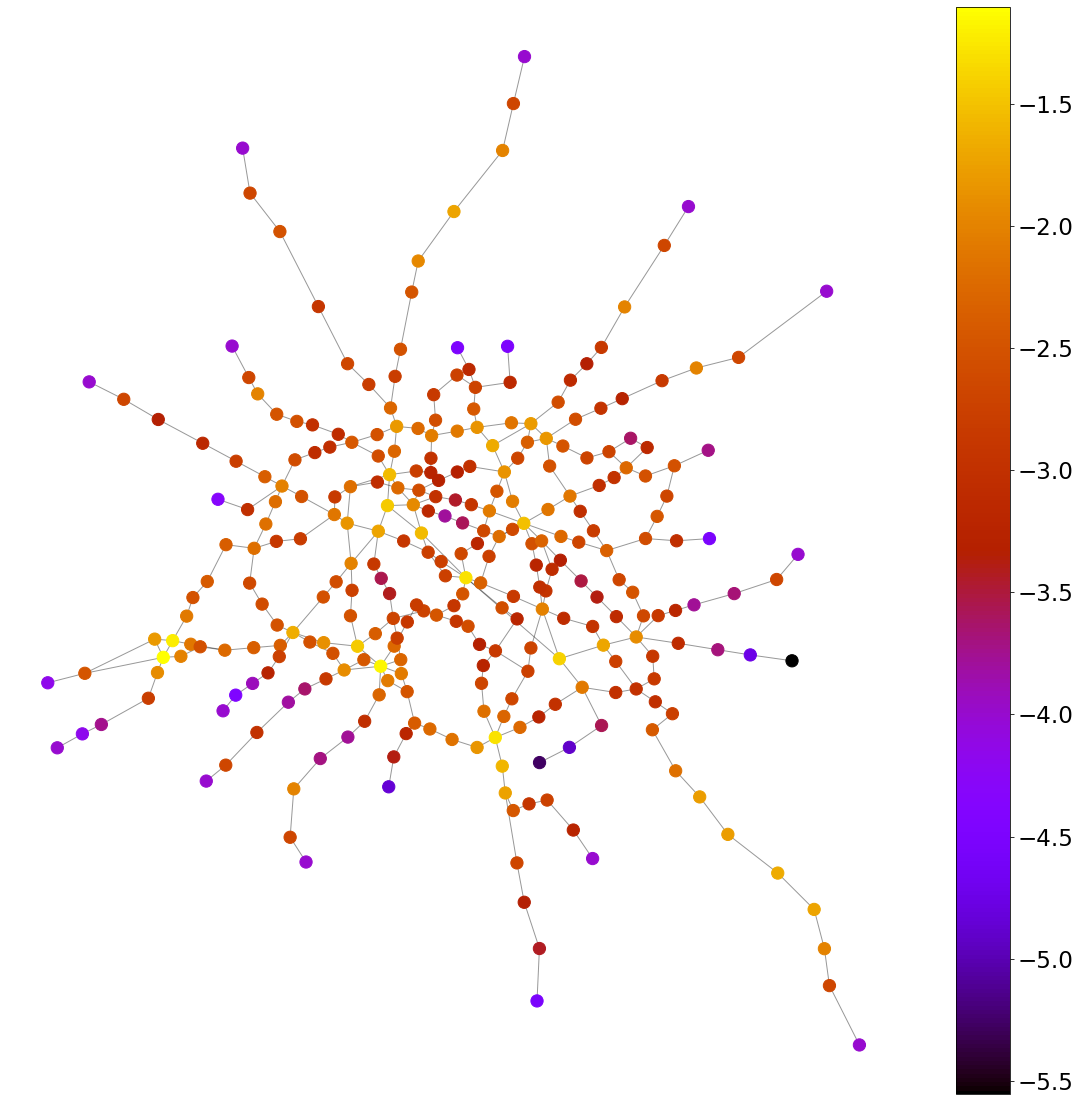}
\includegraphics[scale=0.16]{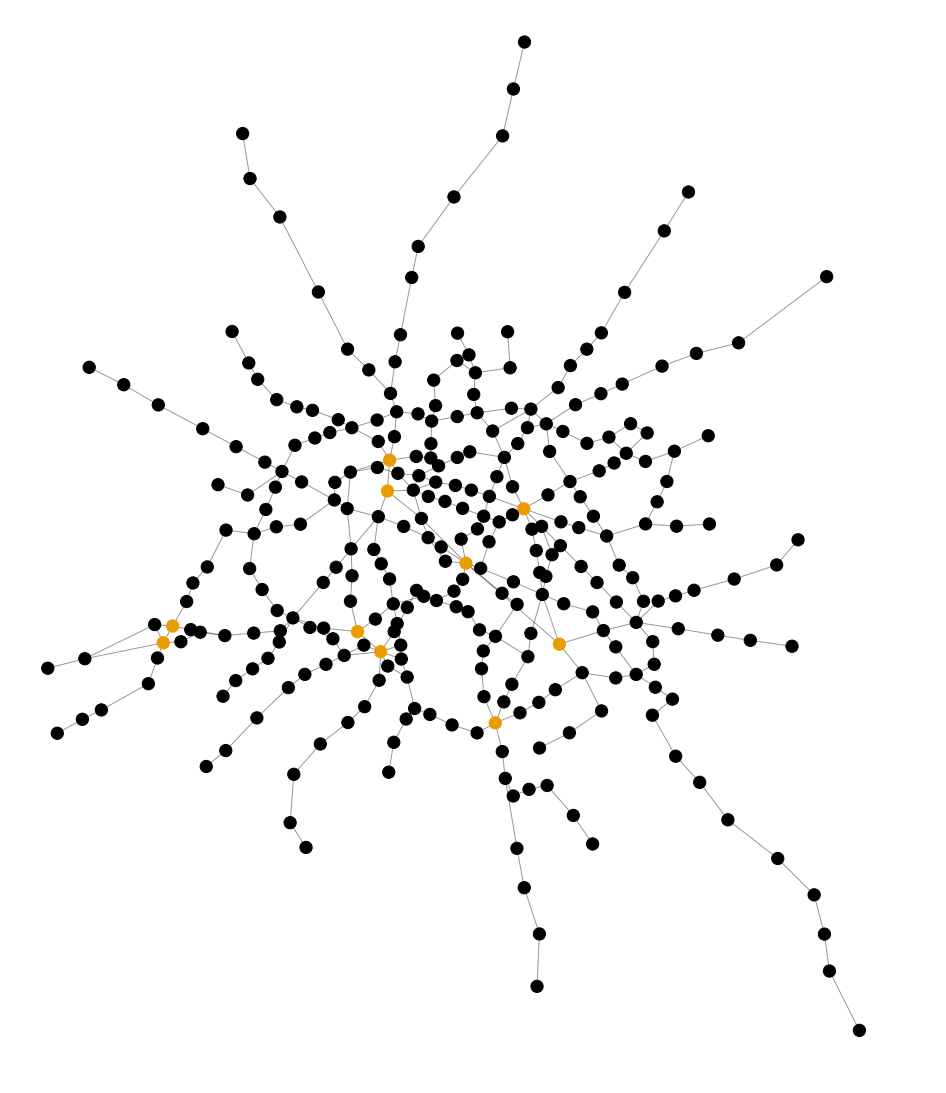}
\includegraphics[scale=0.11]{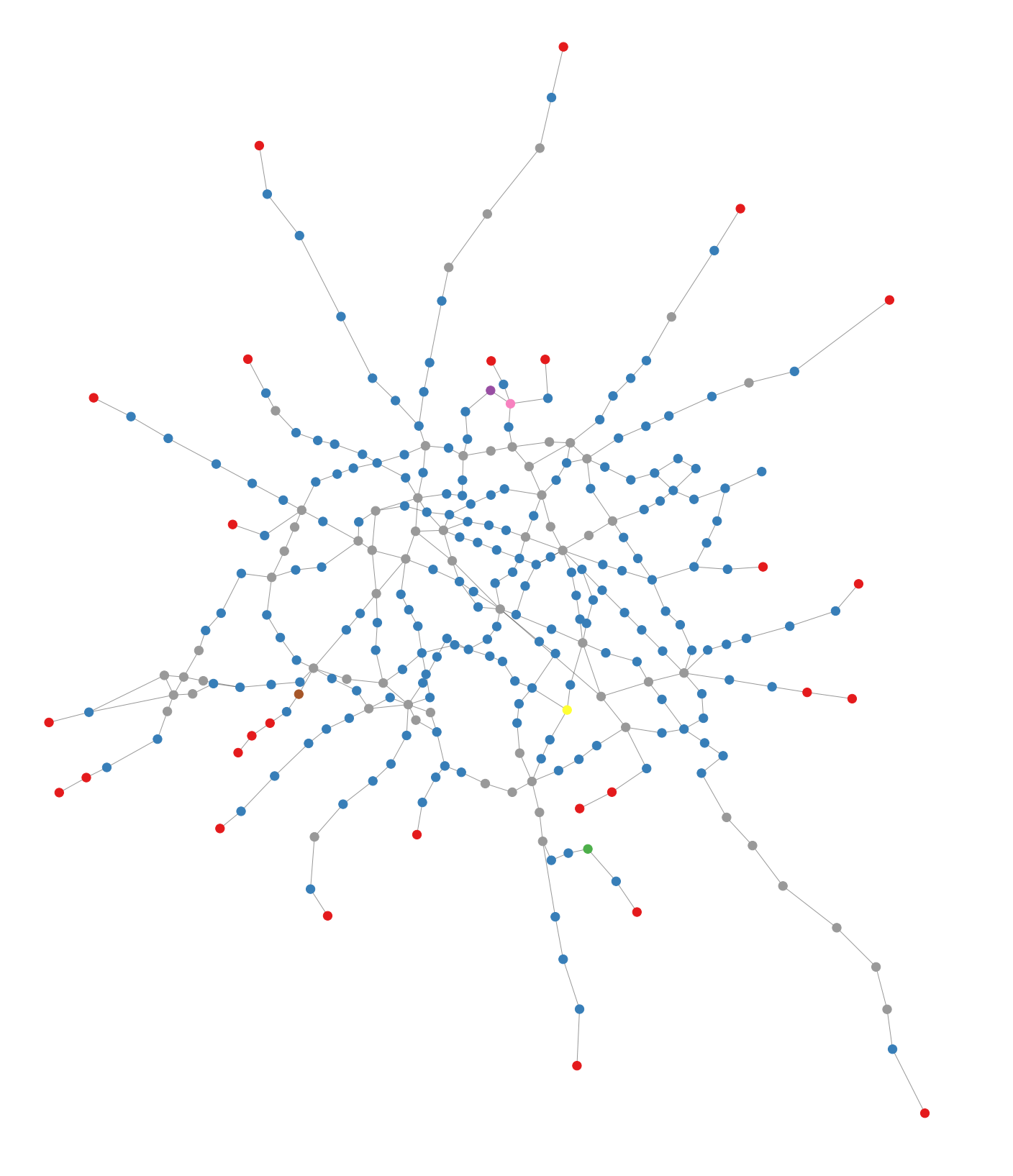}
\caption{Parisian metro system, geographically plotted, results for $l = 5$. The first image shows the value of the $l =5$ Gromov coefficient for the different stops. The second image highlights the 10 stops with the highest $l = 5$ coefficients. The third image shows the results of spectral clustering performed using just the $l = 5$ coefficient. The 10 highlighted stops are: Saint Lazare, Madeleine, Republique, Gare de Lyon, Duroc, Chatelet, Placed'Italie, Montparnasse Bienvenue, Michel Ange Auteuil, Michel Ange Molitor.}
\label{fig:parisl5}
\end{figure}
\begin{figure}[h!]
\centering
\includegraphics[scale=0.16]{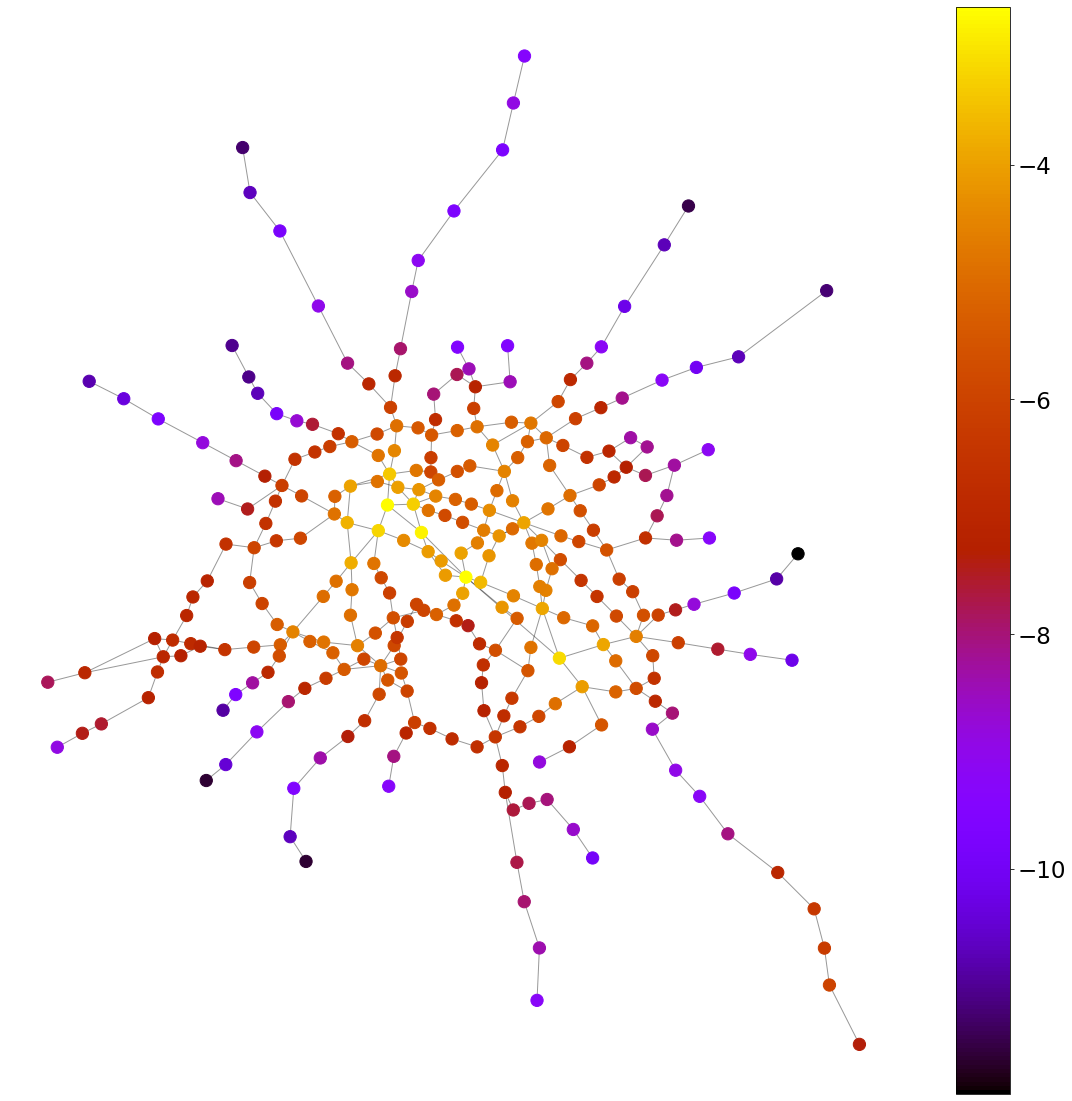}
\includegraphics[scale=0.16]{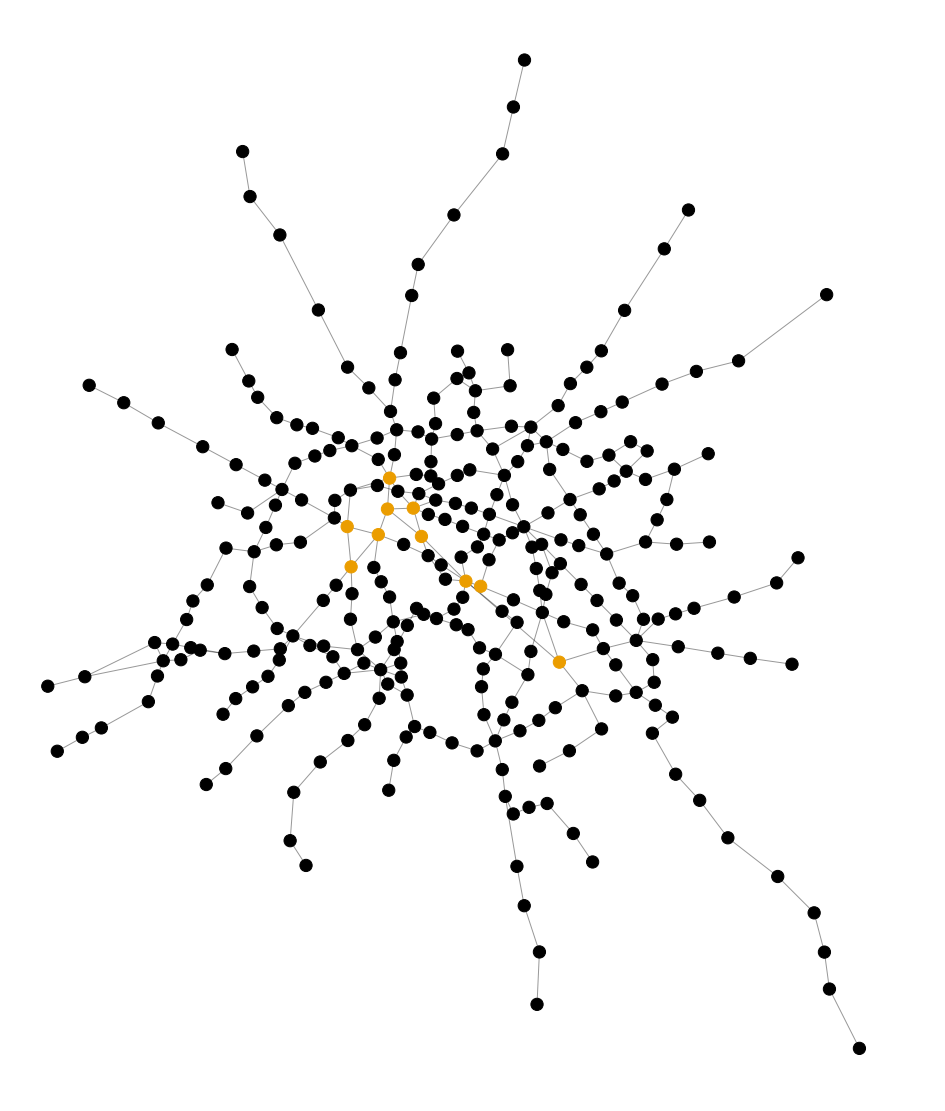}
\includegraphics[scale=0.11]{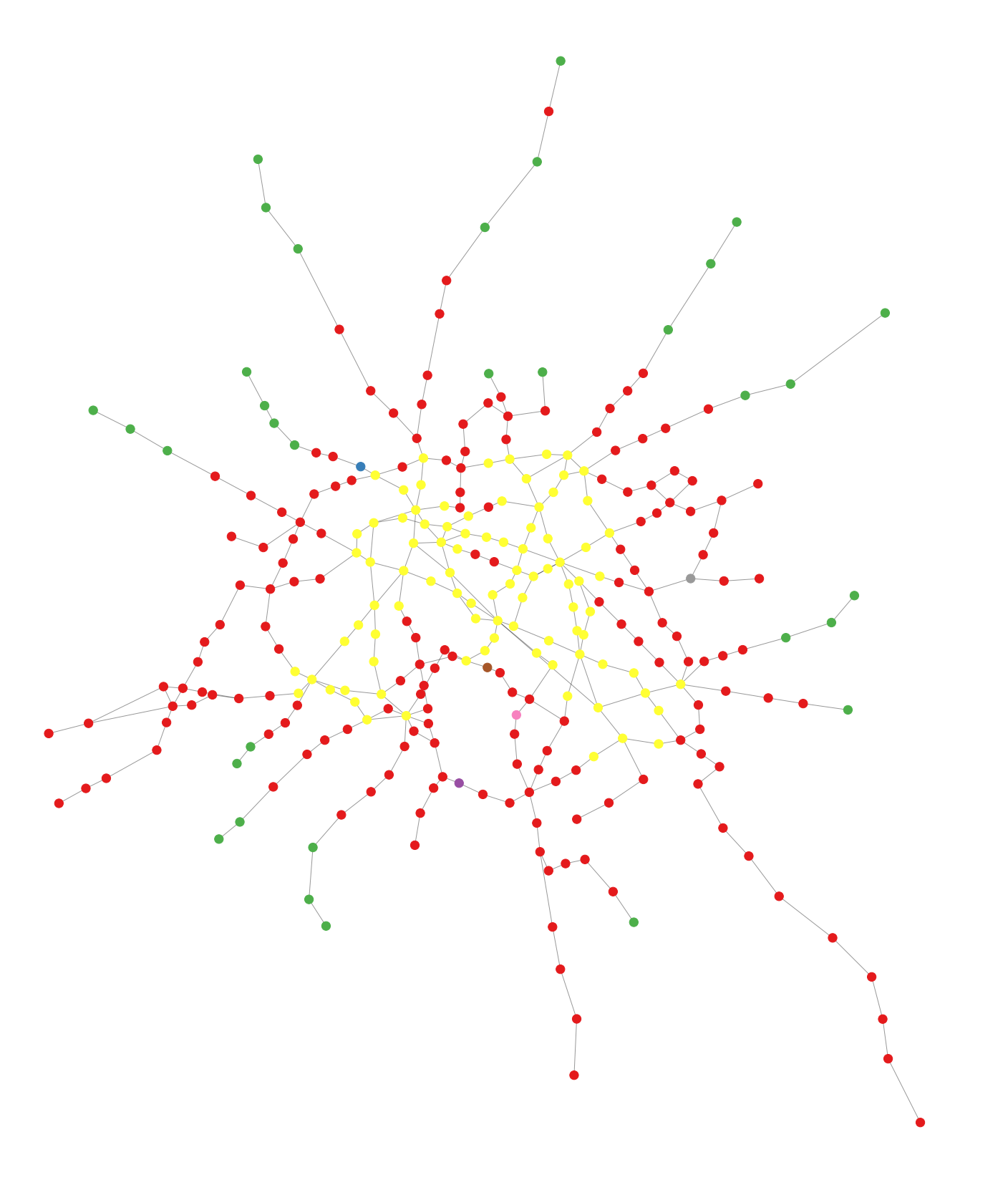}
\caption{Parisian metro system, geographically plotted, results for $l = 10$. The first image shows the value of the Gromov coefficients for the different stops. The second image highlights the 10 stops with the highest coefficients. The third image shows the results of spectral clustering performed using just the $l = 10$ coefficient. The 10 highlighted stops are: Invalides, Champs Elysees Clemenceau, Hotel de Ville, Saint Lazare, Concorde, Gare de Lyon, Madeleine, Opera, Pyramides, Chatelet.}
\label{fig:parisl10}
\end{figure}
\begin{figure}[h!]
\centering
\includegraphics[scale=0.2]{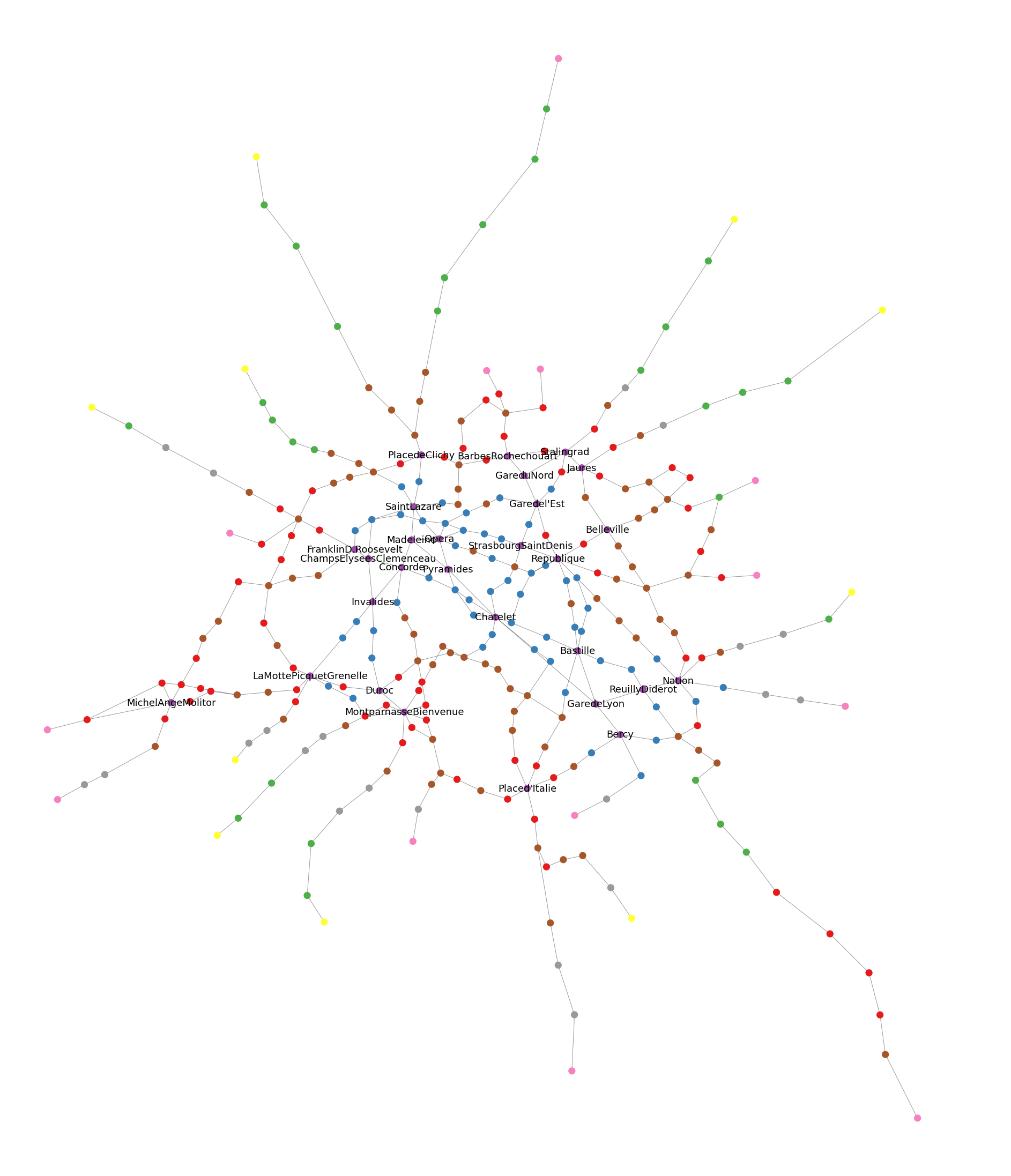}
\caption{Parisian metro system, geographically plotted, clustered using a 3-vector of Gromov coefficients $l = 2, 5, 10$. Shown here with station names labels for the purple cluster.}
\label{fig:3veclabel}
\end{figure}

\begin{figure}[h!]
\centering
\includegraphics[scale=0.16]{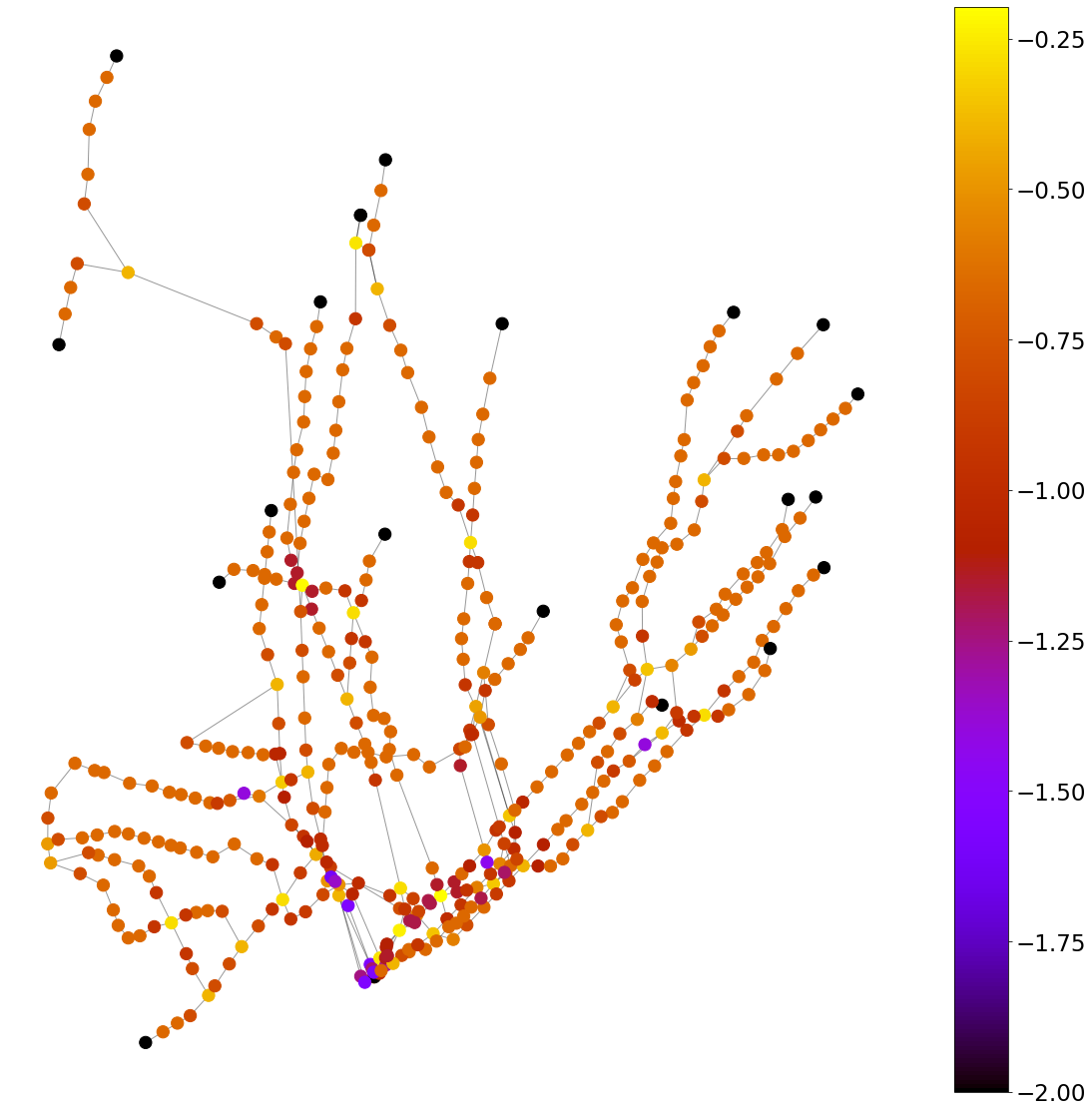}
\includegraphics[scale=0.16]{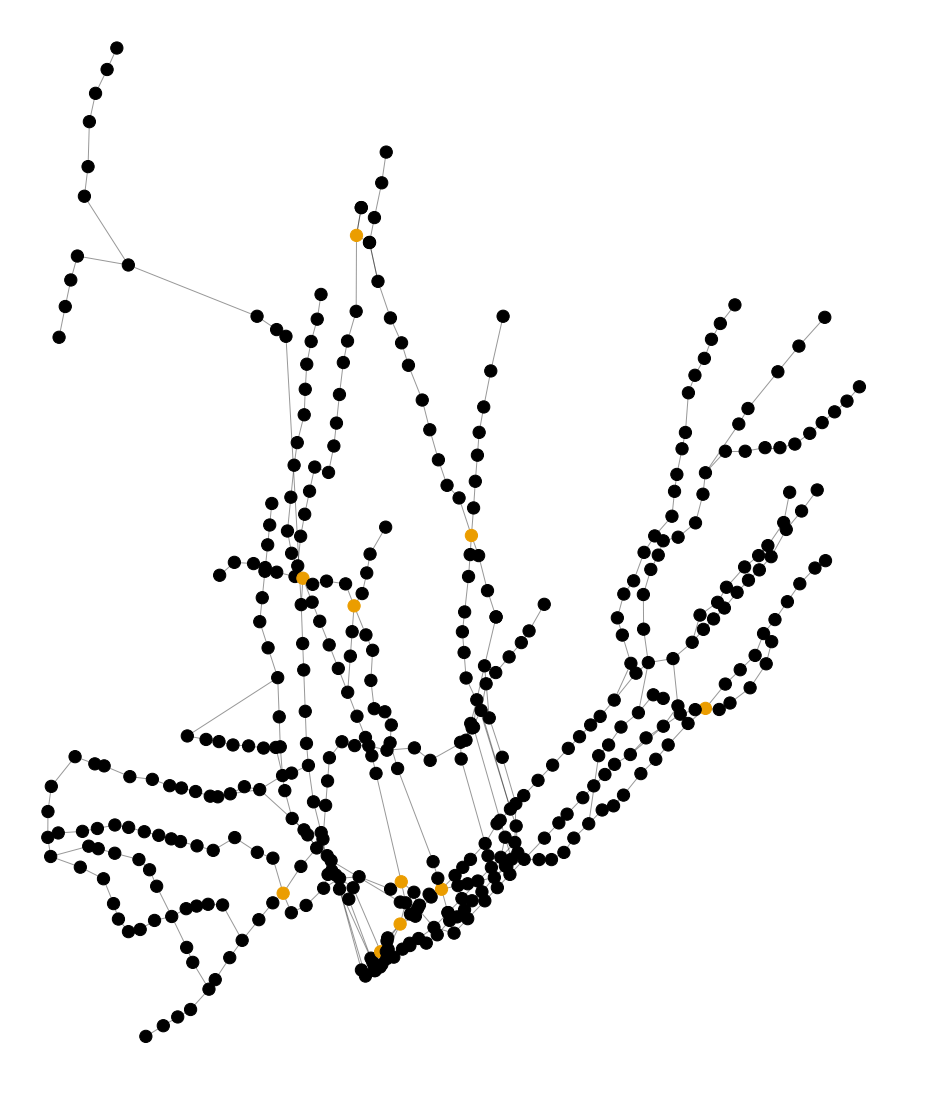}
\includegraphics[scale=0.16]{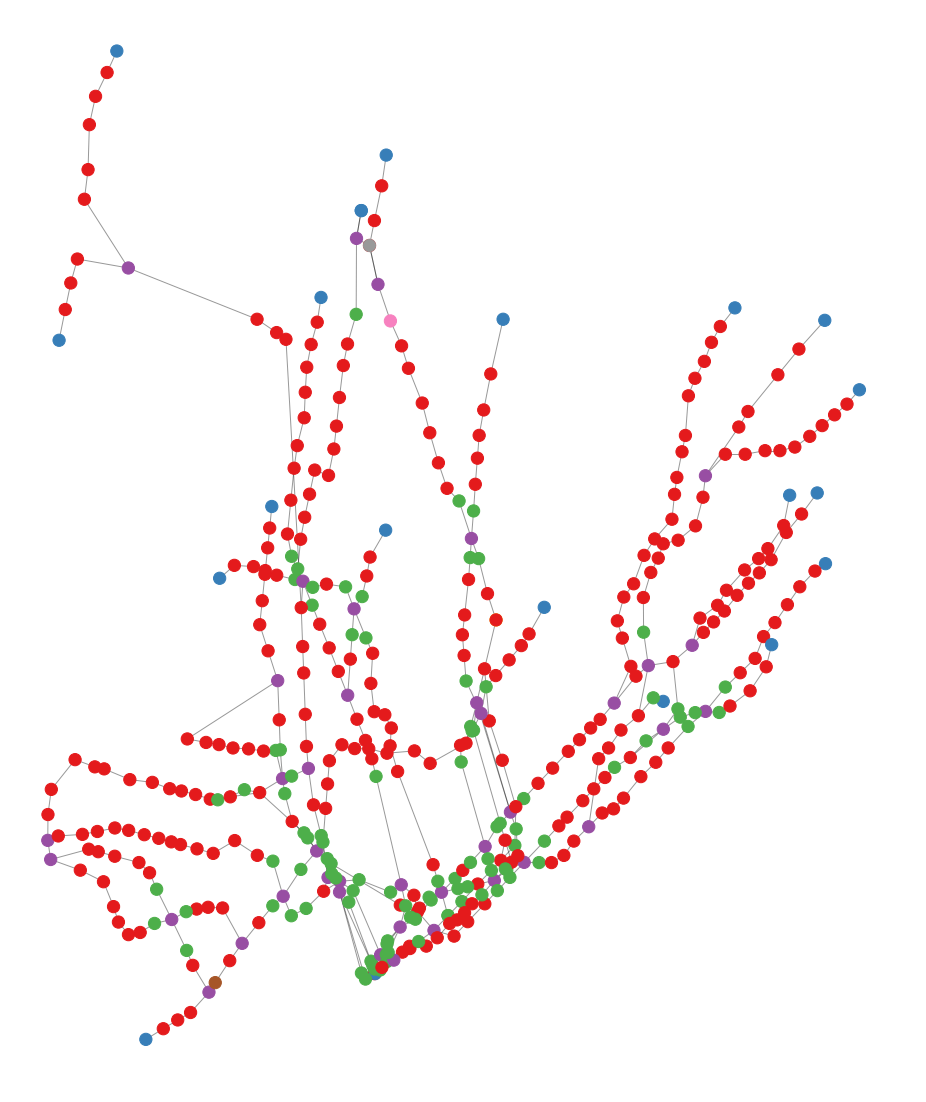}\caption{New York metro system, geographically plotted, results for $l = 2$. The first image shows the value of the $l =2$ Gromov coefficient for the different stops. The second image highlights the 10 stops with the highest $l = 2$ coefficients. The third image shows the results of spectral clustering performed using just the $l = 2$ coefficient. The 10 highlighted stops are: Roosevelt Avenue/74th Street, Fourth Avenue/Ninth Street, Canal Street 46X, 14th Street Union Square, Sutphin Boulevard Archer Avenue JFK Airport, Fulton Street/Broadway Nassau Street, 168th Street, Myrtle Wyckoff Avenues, Delancey Street Essex Street, Broadway Junction.}
\label{fig:nycl2}
\end{figure}

\begin{figure}[h!]
\centering
\includegraphics[scale=0.16]{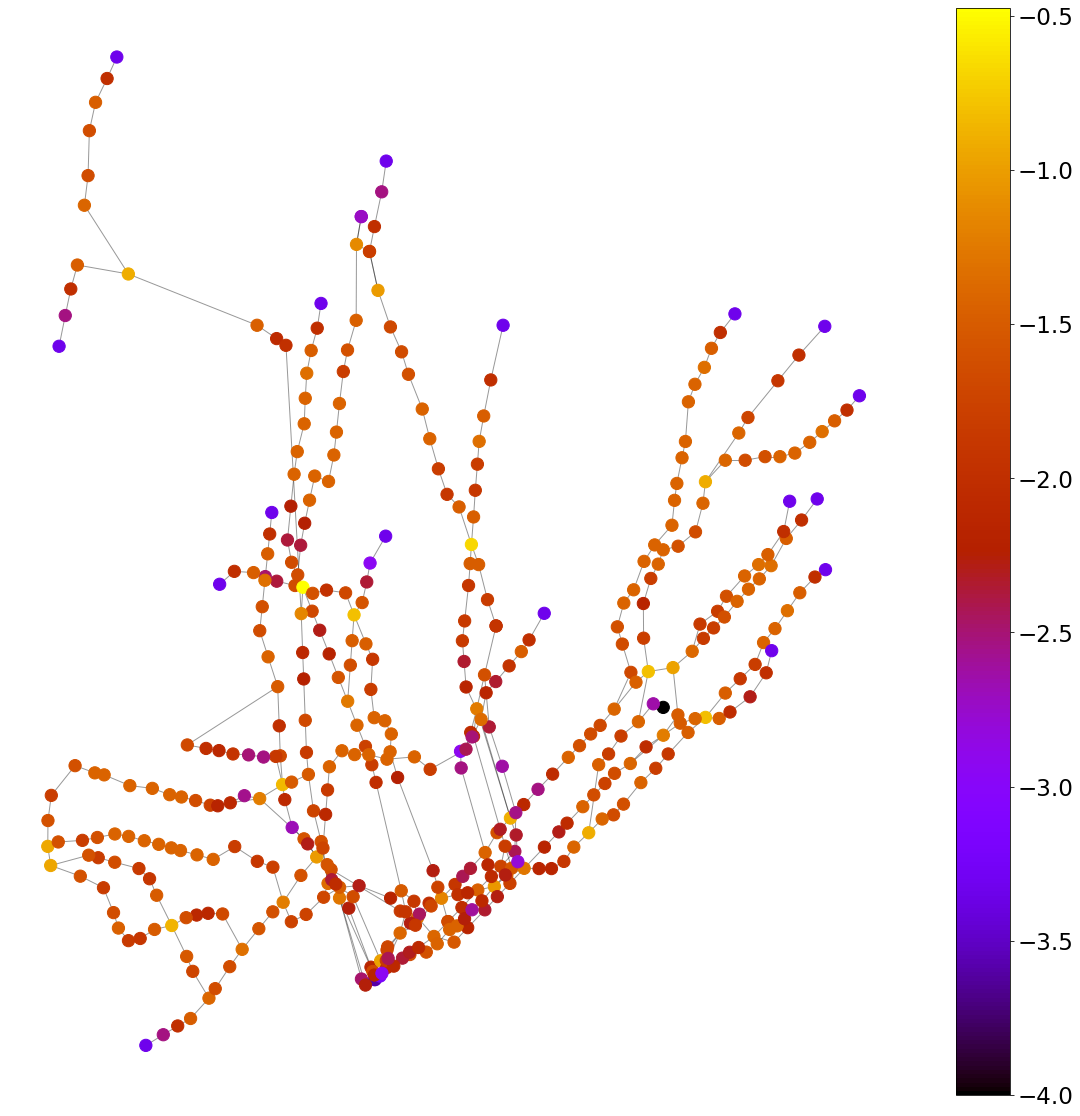}
\includegraphics[scale=0.16]{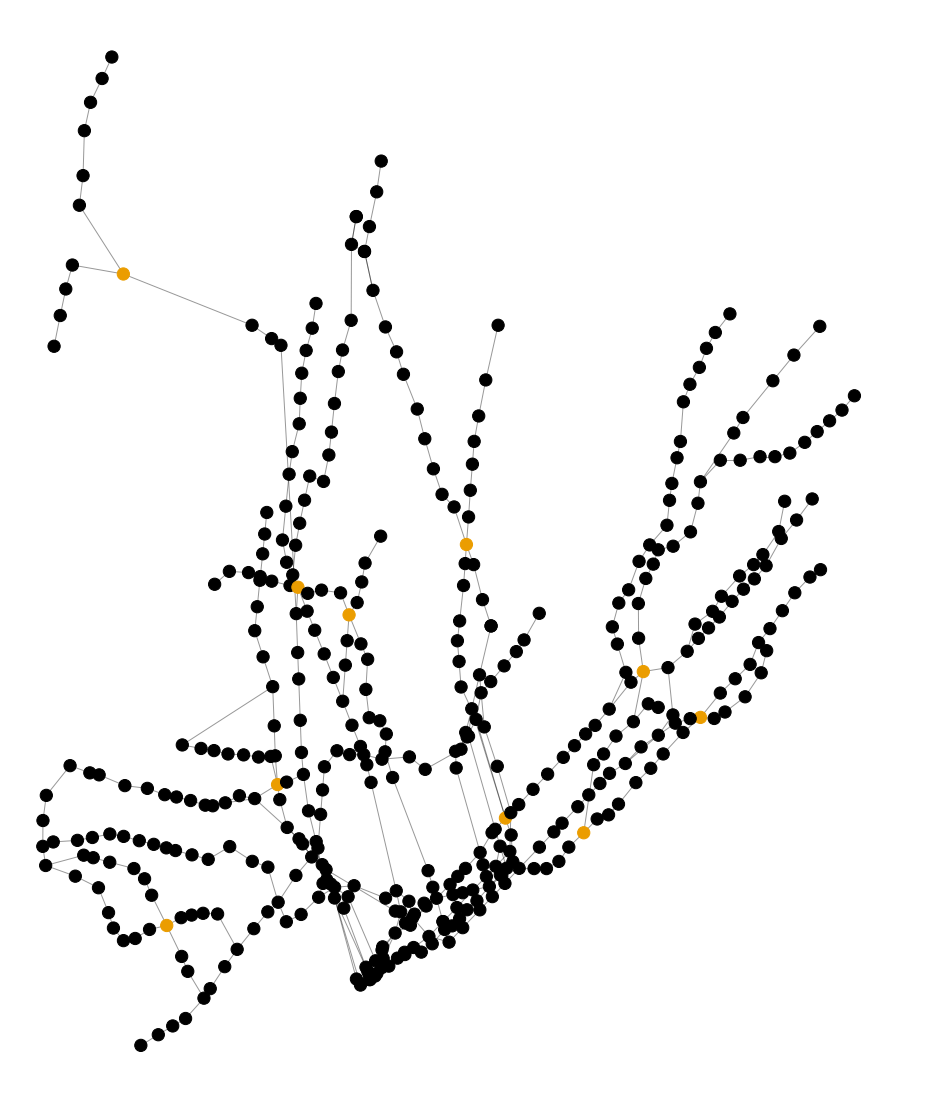}
\includegraphics[scale=0.16]{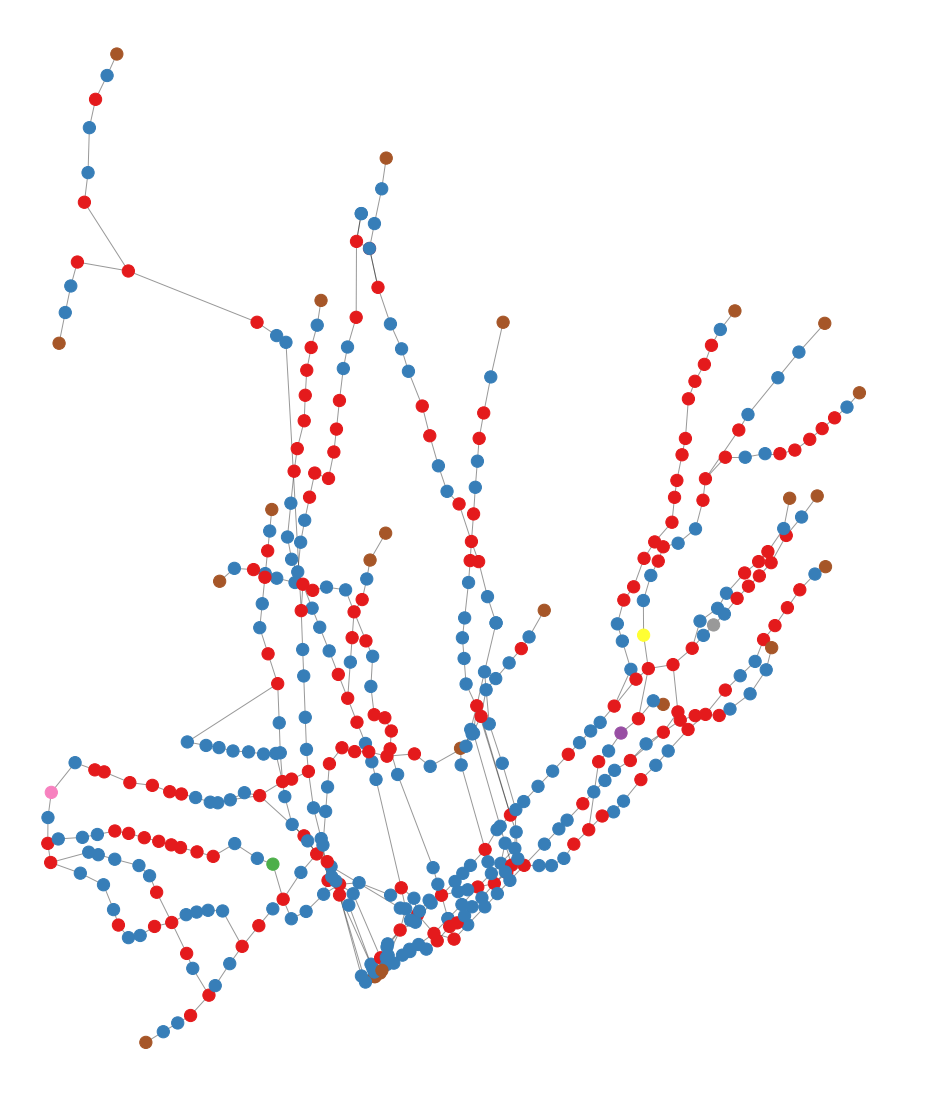}\caption{New York metro system, geographically plotted, results for $l = 4$. The first image shows the value of the $l =4$ Gromov coefficient for the different stops. The second image highlights the 10 stops with the highest $l = 4$ coefficients. The third image shows the results of spectral clustering performed using just the $l = 4$ coefficient. The 10 highlighted stops are: Broad Channel, 96th Street/123, Lexington Avenue/59th Street, New Utrecht Avenue/62nd Street, Franklin Avenue/Botanic Garden, Broadway Junction, 149th Street Grand Concourse, 168th Street, Roosevelt Avenue/74th Street, Myrtle Wyckoff Avenues.}
\label{fig:nycl4}
\end{figure}

\end{document}